\documentclass[aps,pre,twocolumn, longbibliography,superscriptaddress]{revtex4-1}
\usepackage{dcolumn}
\usepackage{bm}
\usepackage{graphicx}
\usepackage{ulem}
 \usepackage{amssymb,wasysym}
 \usepackage{xcolor}

\usepackage[colorlinks=true, allcolors=red]{hyperref}
\DeclareTextSymbol{\degre}{OT1}{23}

\begin{document}
\title{Transition to stress focusing for locally curved sheets}
\author{Thomas Barois, Ilyes Jalisse}
\affiliation{Univ. Bordeaux, CNRS, LOMA, UMR 5798, F-33400 Talence, France}
\author{Lo\"{i}c Tadrist}
\affiliation{Aix Marseille Univ, CNRS, ISM, Marseille, France}
\author{Emmanuel Virot}
\affiliation{hap2U, 75 Avenue Gabriel Péri, 38400 Saint Martin d'Hères, France}
\begin{abstract}
A rectangular thin elastic sheet is deformed by forcing a contact between two points at the middle of its length. A transition to buckling with stress focusing is reported for the sheets sufficiently narrow with a critical width proportional to the sheet length with an exponent 2/3 in the small thickness limit. Additionally, a spring network model is solved to explore the thick sheet limit and to validate the scaling behaviour of the transition in the thin sheet limit. The numerical results reveal that buckling does not exist for the thickest sheets and a stability criterion is established for the buckling of a curved sheet. 
\end{abstract}
\maketitle
\section{Introduction}
The crumpling of a paper sheet is an usual outcome for someone trying to write a deep and meaningful piece of text such as the introduction of a scientific paper. After being crumpled, the paper is irreversibly damaged and its topography may be viewed as a disordered network of ridges.
The persistence of such ridges is a structural memory of stresses focused during the forced manipulation of the sheet. 

Deformations without stretching should prevail for thin sheets due to the large mismatch between thickness and typical size\cite{rayleigh1888bending}. However, stretching deformations appear during crumpling because of stress focusing\cite{witten2007stress}.
Stress focusing is observed with the indentation of a thin plate against a circular contour\cite{cerda1998conical,chaieb1998experimental,cerda1999conical}. The resulting structure is called a developable cone (or d-cone\cite{ben1997crumpled}) and it conveys the idea that the plate adopts a conical shape to satisfy developability.  Experimental d-cones do not rigorously match with conical surfaces. A U-shaped scar is found at the tip of a d-cone and its formation is the consequence of the sheet avoiding diverging curvatures at the apex. This picture of stress focusing can be generalized to crumpling\cite{lobkovsky1995scaling,sultan2006statistics,andresen2007ridge,balankin2010fractal,cambou2011three,croll2019compressive}: a crumpled configuration is
a state occupying a small volume that combines regions approaching developability and localized ridges focusing the stretching deformations. 

In the previous situations reported, a confinement is imposed to a thin sheet and stress focusing is compulsory because there are no developable deformations that can be contained within the imposed volume. 
In this work, we identify the existence of a controlled transition to stress focusing for a thin sheet submitted to a bending deformation and without volume confinement. The main outreach of this work is about the fundamental understanding of crumpling dynamics with the appearance of stress focusing  while smooth solutions of pure bending satisfy the imposed condition. A second outcome is the characterization of the buckling threshold. This has a practical interest for the safe handling of a piece of thin material without inducing stress focusing that would result in structural damages\cite{gottesman2015furrows,chopin2016disclinations}. The characterization of this transition is also of interest in a context of exploiting buckling and crumpling for the realization of functional materials\cite{rogers2009curvy,luo2011compression,mao2012general,ma2012crumpled,yan2013strain,reis2015perspective,holmes2019elasticity}.

The point of this work is to evidence a route to crumpling  via the driving of low-energy  bending  deformations, which can be  viewed  as  a  one  direction  confinement.
 Stress focusing was previously observed for one direction confinement but with peculiar boundary constraints such as an imposed 3-buckle shape\cite{schroll2011elastic} or a pulling force\cite{fuentealba2015transition}. In other situations with frames, the transition to stress focusing can be reversed and defects can appear at small confinement before developable\cite{boudaoud2000dynamics} or smooth\cite{roman2012stress} solutions establish for larger external loads. For draping\cite{cerda2004elements}, a transition to stress focusing exists for a sheet bent under its own weight but the supporting tip adds a localized stress.
\begin{figure}[htbp]
    \centering
    \includegraphics[width=8.5cm]{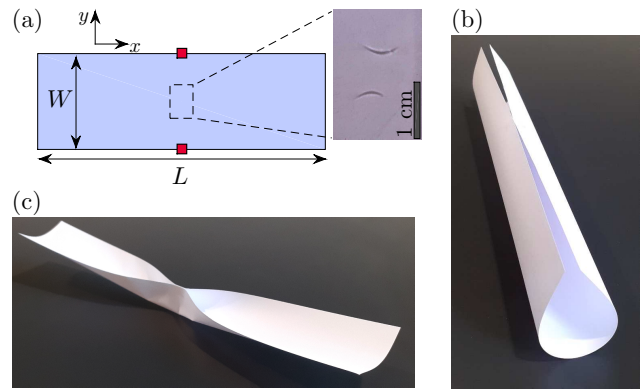}
    \caption{(a) Schematics of a flat sheet of width $W$ and length $L$. The square markers indicate the mid-length points. A picture shows the scars formed after a self-contact procedure with an acetate sheet ($L = 29.7$ cm, $W = 8$ cm, $t= 0.2$ mm) 
    (b) Rectangular paper strip ($L = 29.7$ cm, $W = 15$ cm, $t= 0.1$ mm) with mid-length points at contact and maintained by an adhesive strip. (c) Strip with a smaller width ($L = 29.7$ cm, $W = 5$ cm, $t= 0.1$ mm).}
    \label{fig:1}
\end{figure}

\section{Self-contact experiment}
The present study is based on a simple self-contact procedure applied to a thin sheet in order to characterize the transition to stress focusing. Figure \ref{fig:1} (a) represents an elastic sheet of length $L$ and width $W$. The thickness $t$ is small compared to $W$ and $L$. The material coordinates of the sheet $u,v$ verify $u=x$ and $v=y$ for the flat sheet. The origin for $xyz$ is the center of the sheet. 
Figure \ref{fig:1} shows two elongated sheets for (b) a small ($L/W \approx 2$) and (c) a large aspect ratio ($L/W \approx 6$).

For the aspect ratio $L/W\approx 2$, the sheet is transversely curved without stress focusing. The sheet is however not free of stretching deformation and the presence of a small longitudinal stretching is responsible for the symmetric opening of the sheet. This coupling between transverse bending and longitudinal stretching related to a persistence length for the curved region\cite{lobkovsky1997properties,cerda2003geometry,vandeparre2011wrinkling,barois2014curved,pini2016two,matsumoto2018pinching,taffetani2019limitations}. The limit for a sheet free of stretching would be obtained for a small length $L\ll W$. In this case, the sheet profile would be invariant along $x$ and it would match with Euler's elastica\cite{audoly2010elasticity} model in the projection plane $yz$. This limit identifies a smooth solution with self-contact and free of stretching: for any $L$, a developable surface can be obtained from an elastica profile in the plane $yz$ invariant in the direction $x$.
  
For the large aspect ratio $L/W$, the sheet buckles and scars appear close to the mid-line $u=0$. The picture in figure \ref{fig:1} (a) shows the two scars at the apex of the d-cones formed after self-contact ($L=29.7$ cm, $W=8$ cm and $t=0.2$ mm). For the paper sheet in figure \ref{fig:1} (c) four d-cones are formed. In this case, $L/W$ is larger than for the acetate sheet. This points towards a route to crumpling with an increasing disorder for the sheets of large ratio $L/W$. Here, the experiments are performed close to the transition and the post-buckling morphology does not show the usual ridges network\cite{kramer1997stress} observed at high compaction.

The appearance of stress focusing for a sheet simply curved, even locally, is counterintuitive because the bending modes have an elastic energy orders of magnitude below the stretching modes. For a typical scale $W = 5$ cm for the sheet in figure \ref{fig:1} (c), the ratio bending-to-stretching energy with linear coupling is of the order of $(t/W)^2 \sim 4\times 10^{-6}$. Stress focusing is even more surprising here because bending is a well-known strategy to prevent collapse under external loads, according to the process of curvature-induced rigidity\cite{pini2016two,taffetani2019limitations}.

\section{Scaling law for the buckling transition}
\begin{figure}[htbp]
    \centering
    \includegraphics[width=8.5cm]{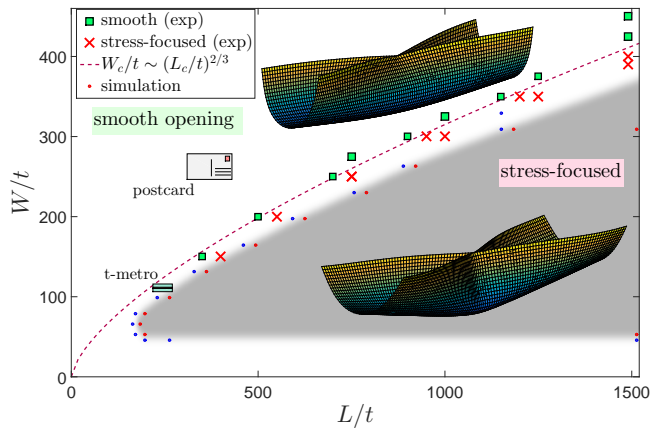}
    \caption{State of a thin sheet in self-contact ($L$: length, $W$: width). The experiments are with acetate sheet (thickness $t = 0.2$ mm) with squares for a smooth opening and crosses for sheets with stress focusing. The dashed line is a power-law with an exponent 2/3 and a numeric prefactor 3.15. The gray area indicates a stress focusing region obtained from the simulation  (blue dots without stress focusing and red dots with stress focusing). A representation of two simulated sheets ($W = 200$) is presented with $L = 500$ for the smooth opening and $L=550$ for the opening with focused stresses. A paper sheet (A4 format) at self-contact remains smooth ($L/t \sim 3000$ and $W/t \sim 2000$, outside the plot range). }
    \label{fig:2}
\end{figure}

The self-contact procedure could be sensitive to unintentional poking or unwanted
gravity load. It is therefore crucial to ensure that these contributions can be
ruled out.
This is done by comparing experimental data to a
numerical model of the self-contact procedure.
A set of experiments is performed with acetate sheets ($t = 0.2$ mm) to identify the buckling transition. Figure \ref{fig:2} presents the state of a sheet at self-contact as a function of $L/t$ and $W/t$. 
 The numerical results are obtained with a spring-network model that will be discussed further. A postcard and a metro ticket are shown to get a sense of the orders of magnitude without the corresponding experiments actually performed. 

For the experiments, buckling occurs in a region with sufficiently large value of $L/t$ for a fixed $W/t$. The smallest width explored is $W = 3$ cm which corresponds to $W/t = 150$. For this value, the bending deformation for self-contact is such that the sheet reaches a plastic regime, which is out of the scope of this work. The largest width explored is $W=9$ cm, slightly above the width for the transition for a sheet of length 29.7 cm ($L/t = 1485$, A4 format). The dashed line is $W_c/t = a (L_c/t)^{2/3}$ with the adjustment $a=3.15$. The experimental transition is reproducible and the existence of a scaling regime validates the idea that such crumpling events relate to a controlled mechanical instability. The value of the exponent 2/3 will be discussed below.

Two strategies were tested to impose the self-contact at  the midpoints ($u=0$, $v\pm W/2$): (i) using two adhesive strips (width 5 mm) and (ii) tightening of a knotted wire (diameter 0.3 mm) between 2 holes close to the midpoints. Both strategies resulted in the same transition which suggest that the detail of the self-contact does not affect the transition. One could indeed wonder if the point-like forces at the self-contact might not play the role of an indenting tip\cite{witten2007stress,cerda1998conical,chaieb1998experimental,cerda1999conical,boudaoud2000dynamics}. This is not the case here because the singularities are always formed away from the self-contacted points.
\section{Spring network model}
\begin{figure}[htbp]
    \centering
    \includegraphics[width=8.5cm]{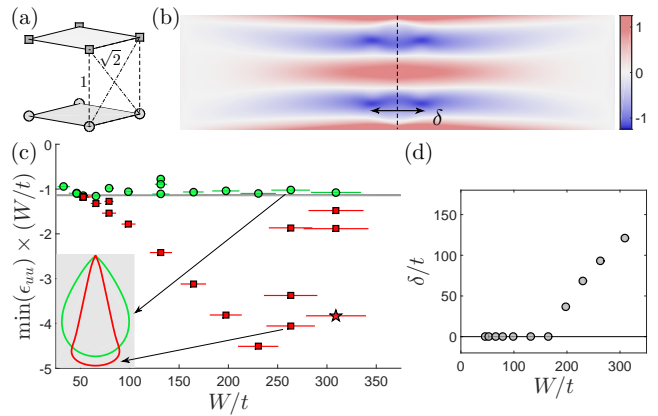}
    \caption{(a) Unit cell of the spring network model. Circles and squares represent the lower and the upper surface. The linear springs of length $1$ and $\sqrt{2}$ are for the first neighbours and the second neighbours  respectively. (b) Normalized longitudinal strain $\epsilon_{uu}\times(W/t)$ of a simulated sheet ($L = 1800$, $W = 470$) just after buckling. The dashed line is the mid-line $u=0$. Positive values corresponds to a stretching and negative values are for a compression. 4 points of stress focusing are present. (c) Normalized minimal strain $\mbox{min}(\epsilon_{uu})(W/t)$ (compression) as a function of the dimensionless width $W/t$. The circles are for simulated sheets with smooth opening and square symbols are for sheets with stress focusing. The horizontal line is at $-1.16$. The profile at mid-length $u=0$ illustrates the curvature sign change with buckling. The star symbol indicates the sheet ($L = 1800$, $W = 470$) selected for the strain profile represented in (b). $t = 1.52$. (d) Horizontal spacing of the local minima for the longitudinal strain. $\delta=0$ means one pair of d-cones at $u=0$ and $\delta>0$ two pairs at $u\pm \delta/2$ as in (b). }
    \label{fig:3}
\end{figure}

For the experimental points below $L/t = 500$, the transition deviates from the scaling regime. To explore the thick sheet regime and the validity of the scaling exponent, a spring network model is computed. Although crystalline spring networks do not rigorously model the isotropic elasticity of continuum media\cite{seung1988defects,curtin1990mechanics,kramer1997stress,ostoja2002lattice,omori2011comparison}, such models are well adapted to explore the scaling properties of thin sheet configurations\cite{didonna2002scaling}.
Figure \ref{fig:3} (a) presents the structure of the unit cell used in the simulation. A thin sheet is obtained by 2D-replication of the unit cell to form a bi-layer. First, a flat sheet is generated and then additional springs are added and tuned to reach self-contact. The equilibrium positions of the vertices are computed by iteration of the spring forces until a static equilibrium is found (see appendix \ref{app:dyn}). 

The spring network model is computed for different sheet dimensions to identify the transition between a smooth opening and a buckled sheet with focused stresses. The criterion to identify a buckled sheet is the sign of the transverse curvature. With a smooth opening, the transverse curvature is always positive. For a buckled sheet, this curvature is partly negative near $u=0$.
Figure \ref{fig:2} represents by dots the simulations performed and a filled area to identify the buckling region. 

Stress focusing is not observed in simulations below $W/t = 56$. For $W/t<56$, the simulations indicate that the sheet can open smoothly for any length $L$. In this regime, there is no risk of damages by stress focusing during manipulation and self-contact. This regime is however difficult to explore with thin elastic sheets. With paper ($t\approx 0.1$ mm), it would require to use strips with $W$ of the order of 5 mm. For such narrow strips, it is not possible to force a self-contact without plasticity. 

The regime of focused stresses is observed for a larger range of parameters in the experiments compared to the simulations. This systematic difference could be explained by the limitations of the spring-network model or the difficulty to realize controlled experiments with thin sheets for which mechanical and physical properties are not perfectly uniform. 

The simulation results  are further processed to find a stability criterion for the transition. Figure \ref{fig:3} (b) represents the longitudinal strain profile $\epsilon_{uu}$ for a sheet of $1800\times 470 \times 2$ vertices ($L/t=1184$, $W/t=309$) with contact of the mid-length points. The strain profile is obtained during the simulated dynamics of the sheet, just after the buckling event. The strain profile combines regions of positive stretching, to allow the sheet to open, and regions of negative stretching (compression). The existence of longitudinal compression induced by transverse curvature was proposed in a previous work\cite{barois2014curved} to satisfy a zero longitudinal net force. Here, we uncover the structure of the stretching strain and the 5-band pattern.  There are four points of maximal compression lying at the middle of the compression bands and longitudinally separated by a distance $\delta$. For a post-buckled sheet, each point of maximal compression is the source point of a d-cone structure. The time-resolved simulation (video, see \cite{video}) clearly shows that the d-cones are not nucleated at the self-contact. This proves that the self-focusing transition is not due to a local indentation by the point-like force at the self-contact.

Figure \ref{fig:3} (c) represents the value of the maximal compression strain $\mbox{min} (\epsilon_{uu})$ for a sheet as a function of the width $W/t$. The strain is normalized with the dimensionless factor $W/t$. Each data point corresponds to a unique couple ($W,L$) and the length $L$ of the sheets is indicated by a horizontal bar for each data point. The plot shows two families of data points: the circle scatters are for the smooth sheets, without curvature inversion, and the square or star scatters are for the post-buckled sheet with curvature inversion of the transverse profile (see inserted schematics). The data point with a star corresponds to the dimension for the strain map in figure (b). 

The plot in figure \ref{fig:3} (c) indicates a threshold value  $\mbox{min} (\epsilon_{uu})\times(W/t) = -1.16$ below which a sheet at self-contact buckles. The buckling dynamics can be followed in more details with the successive profiles of the sheet during the simulation. Figure \ref{fig:s11_minstrain} in the appendix represents the evolution of the maximal compression strain during the convergence of the mechanical equilibrium of the simulated sheet. The sheet with $1700\times 470 \times 2$ vertices ($L/t=1118$, $W/t=309$) converges to a smooth solution with a minimal strain above the threshold $-1.16$. For the simulation with 1750 and 1800 vertices ($L/t=1151$ and $L/t=1184$), the minimal strain is sufficiently low to cross over the threshold value -1.16 and a sudden drop of the minimal strain is observed as a signature of stress focusing. For the small sheet with $W/t =35$ and $W/t =49$, there is no buckling transition because the compression strain remains below the threshold $1.16\times (t/W)$ for any $L/t$.

Figure \ref{fig:3} (d) represents the longitudinal spacing $\delta$ between the points of minimum strain. A value $\delta = 0$ means that there is a total of 2 local minima for the compression strain located on the mid-length line $u = 0$. With $\delta>0$, there are 4 local minima for the compression strain as shown in figure \ref{fig:3} (b). The number of local minima sets the profile of the first buckling event. 
In the simulation with $1750 \times 470 \times 2$ vertices, the stress focusing is first triggered at the 4 symmetric points of maximal compression strain. Then, the 4 points travel towards the mid-length line and an abrupt increase of the maximal compression is observed when the pairs of stress focusing points merge.

\section{Analytical toy model}
The transition to stress focusing identified in experiments and simulations can be rationalized by a toy model based on (i) elastic energy minimization \cite{barois2014curved} and (ii) a buckling criterion \cite{hutchinson2010knockdown}. 

To estimate the elastic energy, we assume that the sheet at self-contact is described by its typical transverse curvature that takes the value $c_0\sim 1/W$ at $u=0$ and $c_e$ at the free ends at $u=\pm L/2$. By considering only the first-order contributions, the total elastic energy scales as:
\begin{equation}
    \frac{\mathcal{E}(c_e)}{ELWt} \sim t^2({c_0}^2 + {c_e}^2) + t^2\frac{W^4}{L^4}(c_0-c_e)^2 +  \frac{W^8 (c_0-c_e)^4}{L^4}\label{eq:mod}
\end{equation}
with $E$ the Young modulus. The terms in the right hand-side stand for the transverse bending energy, the longitudinal bending energy and the longitudinal stretching energy (see appendix \ref{app:energy}). $c_e$ is the only free parameter of the model and its value is obtained by the minimization of the energy $\partial \mathcal{E}(c_e)/\partial c_e = 0$ that yields to an equilibrium between two bending terms and one stretching term:
\begin{equation}
    t^2 c_e - t^2\frac{W^4}{L^4}(c_0-c_e) = 2\frac{W^8(c_0-c_e)^3}{L^4}
    \label{eq:min_E}
\end{equation}
 We discuss here the solution of equation (\ref{eq:min_E}) in the thin sheet limit: with $W \gg t$ and $L \gg t$, the transition is such that $L$ is larger than $W$ and we can assume $L^4 \gg W^4$. Consequently, the term $t^2 c_e \sim t^2/W$ dominates the left hand-side in equation (\ref{eq:min_E}) and the scaling for the curvature difference is $c_0 - c_e \sim t^{2/3}L^{4/3}/W^3$. According to our previous work\cite{barois2014curved}, the scaling for the compression induced by a transverse bending is $\epsilon_{uu} \sim W^4 (c_0-c_e)^2/L^2$. Considering the criterion $|\epsilon_{uu}|(W_c/t) \sim  1$, we obtain 
\begin{equation}
W_c \sim t^{1/3}{L_c}^{2/3}
\label{eq:scaling}
\end{equation}

The exact solution of equation (\ref{eq:min_E}) confirms that the scaling (\ref{eq:scaling}) is valid only for large $W/t$ (see appendix \ref{app:energy} and figure \ref{fig:phase_diag}). This exact solution also predicts an absence of buckling transition for small $W/t$ as found in the simulations. 

Experiments were also performed with paper and the transition was found for lower $L/t$ (not shown in figure \ref{fig:2} for clarity, see appendix \ref{app:exp_pap}). This difference is attributed to the anisotropy of paper \cite{alava2006physics}. By taking into account the anisotropy \cite{matsumoto2018pinching} with a smaller Young modulus for the transverse direction, we show in the appendix \ref{app:exp_pap}  that the prefactor for the scaling relation is modified accordingly.

\section{Conclusion}
The present results have evidenced a simple scaling relationship for the critical compressive strain of a sheet buckling with stress focusing, that is $\epsilon_c \sim t/W$ with a prefactor close to unity, as shown in figure \ref{fig:3}(c). This criterion is used in the toy model and the state diagram obtained from the exact solution of equation (\ref{eq:min_E}) is consistent with experiments and simulations. In our previous work\cite{barois2014curved}, the existence of a longitudinal compression for a transversely curved sheet was proposed to satisfy internal mechanical equilibrium in a model. 
Its relevance was only justified indirectly via a detailed analysis of the persistence length. 
Here, we show that the induced compression strain has the same effect than an external compression load applied to a shell with the same curvature:
the geometric stability criterion $\epsilon_c \sim t/W$, with $1/W$ the typical  curvature of the transversely curved sheet, is the equivalent of the classical buckling of perfect cylindrical shells, where linear stability analysis predicts $\epsilon_c = 1/\sqrt{3(1-\nu^2)}\times t/R$ \cite{hutchinson2010knockdown}, with $\nu$ denoting the Poisson coefficient of the shell and $1/R$ its curvature.
Notably, the above-mentioned formula infamously overestimates the buckling load of cylindrical shells \cite{peterson1968buckling}, because those systems are extremely imperfection-sensitive \cite{singer2002vol, gerasimidis2018establishing}. 
This affects the reliability of thin shell structures, even for precisely manufactured shells such as soda cans \cite{virot2017stability} and space rockets \cite{peterson1968buckling}.
In contrast, the present work identifies a strain threshold $\epsilon_c$ that leads to robust scaling regimes observed in experiments and simulations.
This hints that buckling by self-contact of a curved sheet is relatively insensitive to imperfections. We attribute this to a buckling threshold reached locally, contrarily to cans for which buckling can be triggered at any weak spot.
Furthermore, while the buckling of cans by external loads leads to collapsing, buckling by self-contact offers more control with a self-regulation of the internal stress after the transition.
This provides an interesting strategy to  reliably  control  buckling  events  in  functional and  smart  materials
by choosing geometries close to the transition to take advantage of the fast morphology switch. 
\appendix
\section{Toy model and theoretical state diagram}\label{app:energy}
\begin{figure}[htbp]
    \centering
    \includegraphics[width = 8.5cm]{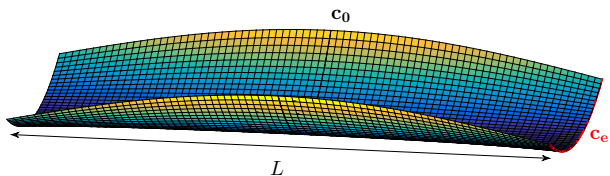}
    \caption{View of strip of width $W$ with the shape of a surface ruled by the curvature function $c(u)$ with $-L/2 < u < L/2$.  }
    \label{fig:toy_model}
\end{figure}

The sheet surface can be described by a curvature function $c(u)$ (see figure \ref{fig:toy_model}). For a simple scaling approach, we reduce the sheet kinematics to 2 degrees of freedom for the transverse curvature:
\begin{itemize}
    \item[-]  $c_0$ for the curvature at mid-length, and
    \item[-]  $c_e$ for the curvature at the free end $u=\pm L/2$.
\end{itemize}

To estimate the elastic energy of the sheet, we will consider the first order contributions from the quadratic variables obtained from the 2 degrees of freedom $c_0$ and $c_e$.

For the transverse bending energy, we assume that the energy comes from the two contributions:
\begin{equation}
\mathcal{E}_{b,\perp} \sim \alpha ELWt^3 {c_0}^2 + \beta ELWt^3 {c_e}^2 
\label{eq:bt}
 \end{equation}
in which $E$ is the young modulus, $L\times W \times t$ the physical dimensions of the sheet and $\alpha,\beta$ two numeric prefactors that could be determined with a more detailed analysis of the sheet's state. In the following, the numerical prefactors will be omitted, given that we are discussing  scaling relations here.

To estimate the longitudinal bending energy, we use the relation that gives the vertical displacement of the sheet 
\begin{equation}
Z(u,v) = \frac{1}{2}c(u)v^2
\end{equation}
that gives the scaling for the typical vertical displacement
$
Z_0 \sim c_0 W^2
$
and
$
Z_e \sim c_e W^2
$.
In a first approximation, the longitudinal curvature is 
\begin{equation}
\frac{1}{R_{long}} \sim \frac{Z_0 - Z_e}{L^2}
\end{equation}
which gives an estimation for the longitudinal bending energy
\begin{equation}
\mathcal{E}_{b,\parallel} \sim  ELWt^3 \frac{W^4}{L^4} (c_0 - c_e)^2
\label{eq:bl}
\end{equation}

Finally, the longitudinal stretching strain is obtained by the ratio 
\begin{equation}
\epsilon = \frac{\sqrt{L^2 + (Z_0-Z_e)^2} - L}{L^2}\label{eq:strain}
\end{equation}
in which the square root estimates the excess of length at the edge of the sheet because of the relative vertical displacement $|Z_0 - Z_e|$. After a Taylor expansion for $Z_0 - Z_e \ll L$, the stretching energy is obtained with
\begin{equation}
\mathcal{E}_{s,\parallel} \sim ELWt \frac{W^4}{L^4} (c_0 - c_e)^4
\label{eq:sl}
\end{equation}

Using equations (\ref{eq:bt}), (\ref{eq:bl}), and (\ref{eq:sl}), we get the scaling for the total energy used in the article:
\begin{equation}
\mathcal{E} \sim  \mathcal{E}_{b,\perp} +\mathcal{E}_{b,\parallel} + \mathcal{E}_{s,\parallel}\label{eq:Etot}
\end{equation}

In figure \ref{fig:toy_model}, the sheet is represented for fairly low curvatures $c_0$ and $c_e$ in order to visualize easily the sheet's shape. For self-contact, the curvature at mid-length scales as
$
c_0 \sim 1/W
$.

With the self-contact condition, the only free parameter in equation (\ref{fig:toy_model}) is $c_e$. The local optimal for the energy is obtained from the cancellation of the derivative with respect of $c_e$ which reads:
\begin{equation}
t^2{c_e} - t^2 \frac{W^4}{L^4}(c_0-c_e) - 2\frac{W^8}{L^4}(c_0 - c_e)^3 = 0  \label{eq:E_min}
\end{equation}

With the variable $\delta_c = c_0-c_e$, we get
\begin{equation}
2\frac{W^8}{L^4}{\delta_c}^3 + t^2\left(1 + \frac{W^4}{L^4}\right)\delta_c -t^2{c_0}  = 0  \label{eq:E_card}
\end{equation}

Equation (\ref{eq:E_card}) is a third-order polynomial that may be written $\delta_c^3 + p \delta_c + q = 0 $ to fits Cardano's formula with 
\begin{eqnarray}
p &=& \frac{L^4t^2}{2W^8}\left(1 + \frac{W^4}{L^4}\right)\\ 
q &=&-\frac{L^4t^2}{2W^9}
\end{eqnarray}
with the condition $c_0 = 1/W$. The solution of equation (\ref{eq:E_card}) is then:
$$
\delta_c = \sqrt[3]{-\frac{q}{2} + \sqrt{\frac{q^2}{4}+\frac{p^3}{27}}}+\sqrt[3]{-\frac{q}{2} - \sqrt{\frac{q^2}{4}+\frac{p^3}{27}}}\label{eq:sol}
$$

The stretching energy involves region of positive (extension) and negative (compression) strain  with the same scaling magnitude
$$
\epsilon = \frac{W^4{\delta_c}^2}{L^2}
$$
obtained after the Taylor expansion of equation (\ref{eq:strain}).

\begin{figure}[htbp]
    \centering
    \includegraphics[width=8.5cm]{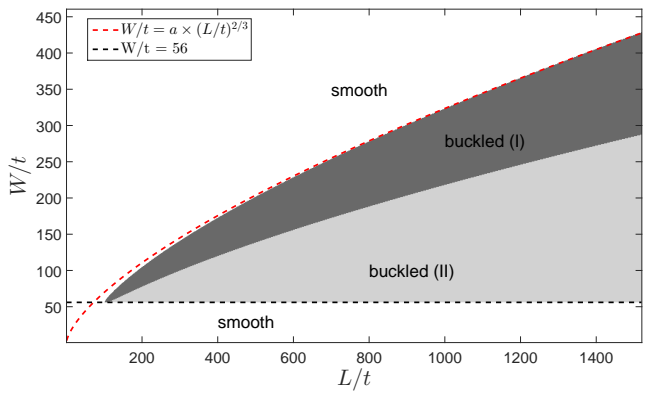}
    \caption{Buckling criterion $\epsilon < \epsilon_c$ (eq. (\ref{eq:crit})). In the region \textit{buckled I}, the buckling criterion is satisfied. Elsewhere, the sheet should be smooth except in region \textit{buckled II} (see text for details). The solid line is the scaling law with exponent 2/3 and a numerical factor $a = 3.235$.  }
    \label{fig:phase_diag}
\end{figure}

If we use a stability criterion  found in the simulations and consistent with the buckling of cans under axial load:
\begin{equation}
\epsilon < \epsilon_c\mbox{, with } \epsilon_c = \alpha \frac{t}{W}
\label{eq:crit}
\end{equation}

with $\alpha = -0.43$, we find the phase diagram presented in figure \ref{fig:phase_diag}.

The diagram $(L/t, W/t)$ shows a dark filled area (buckled I) for which the buckling criteria (\ref{eq:crit}) is satisfied. The buckled region \textit{buckled (I)} has a tongue-like structure. Away from this region, the sheet's shape should be smooth. This is the case almost everywhere in experiments and simulations, except in the region identified by \textit{buckled II}. Although the model does not predict stress focusing for this region, we suspect that the sheet should be buckled here as well: for each point in the region \textit{buckled II}, one can find a point for the same $W/t$ and a smaller length $L/t$ in the effectively buckled region \textit{buckled I}. Formulated the other way around: from a sheet effectively buckled in region  \textit{buckled (I)}, the sheet should remain buckled if it was longer (larger $L$ for same $W$). We think that the absence of buckling in region \textit{buckled (II)} is a flaw of our model that oversimplifies the curvature decay profiles and how $L$ affects the scaling of the energy terms in equation (\ref{eq:Etot}).   

This model captures the thick-sheet regime and the fact that buckling does not exist below a threshold value $W_c/t$ identified by a dashed line in the figure. The value of the prefactor $\alpha=-0.43$ sets the value of the lower bound for the transition $W_c/t = 56$ and the value of the prefactor $a$. 

An additional red solid line is added for the transition regime in the thin-sheet limit $W/t = a\times (L/t)^{2/3}$ with $a = 3.235$. This regime consistently matches with the region \textit{buckled I} for $W/t$ and $L/t$ sufficiently large. This regime can be found from the model if the term for the longitudinal stretching is neglected (see main article for the scaling argument with $L^4\gg W^4$). This also justifies that for thick sheets, both numerical and experimental data points should shift towards smaller values of $W/t$. 
 \section{Experiments with paper}\label{app:exp_pap}
\begin{figure}[htbp]
    \centering
    \includegraphics[width=8.5cm]{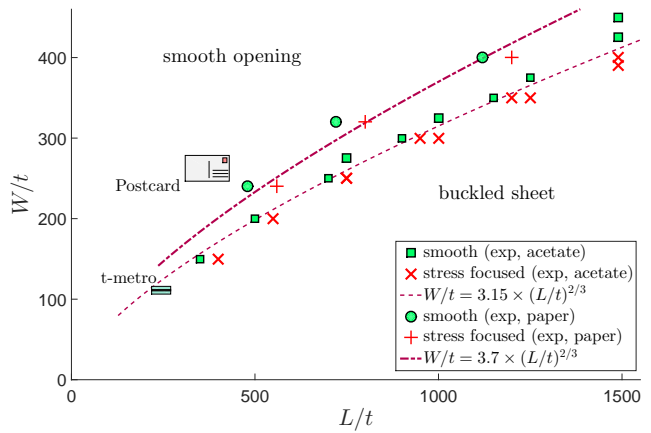}
    \caption{State of rectangular sheets of dimensions $L,W,t$ with contact of the mid-length points. The plot represents experiments with acetate sheets and paper. The scaling law $W/t \sim (L/t)^{2/3}$ is represented with two different numerical prefactors.}
    \label{fig:s1}
\end{figure}

 The self-contact experiment was performed with acetate sheets (see figure 1 of the main article) and standard paper sheets of thickness $t = 0.125$ mm. Figure \ref{fig:s1} presents the state of a rectangular sheet curved by self-contact for acetate sheets (squares and crosses (x)) and paper (circles and crosses (+)).

\begin{figure}[htbp]
    \centering
    \includegraphics[width=8.5cm]{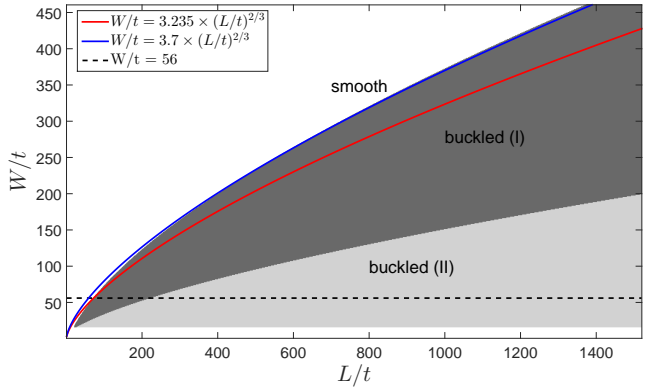}
    \caption{State diagram predicted by the toy model with anisotropy (equation (\ref{eq:Etot_anis}), $E_\perp/E_\parallel = 0.71$). The dashed line and the solid red line (with factor 3.235) are the same as in figure \ref{fig:phase_diag} for the isotropic model $E_\parallel = E_\perp$. The solid blue line is the scaling law with a prefactor 3.7 in good agreement with the experiments performed with paper sheets.  }
    \label{fig:state_diag_anis}
\end{figure}

Paper sheets buckle more easily than acetate sheets: in the region between the two scaling relations $W/t = 3.15 (L/t)^{2/3}$ and $W/t = 3.7 (L/t)^{2/3}$ in figure \ref{fig:s1}, paper sheets are buckled but acetate sheets open smoothly. This difference can be rationalized by introducing anisotropy and two different Young's moduli in the equation (\ref{eq:Etot}) for the energy 
\begin{equation}
\tilde{\mathcal{E}} \sim  \tilde{\mathcal{E}}_{b,\perp} +\tilde{\mathcal{E}}_{b,\parallel} + \tilde{\mathcal{E}}_{s,\parallel}\label{eq:Etot_anis}
\end{equation}
in which 
\begin{equation}
    \tilde{\mathcal{E}}_{b,\perp} \sim LWt \times E_\perp t^2({c_0}^2 + {c_e}^2) 
\end{equation}
is the transverse bending energy,
\begin{equation}
    \tilde{\mathcal{E}}_{b,\parallel} \sim LWt \times E_\parallel t^2 \frac{W^4}{L^4}(c_0-c_e)^2 
\end{equation}
is the longitudinal bending energy and
\begin{equation}
    \tilde{\mathcal{E}}_{s,\parallel} \sim LWt \times E_\parallel \frac{W^8}{L^4}(c_0 - c_e)^4 
\end{equation}
is the longitudinal stretching energy $E_\parallel$ and $E_\perp$ are two Young's moduli  with $E_\perp < E_\parallel$ in which $\parallel$ refers to the direction parallel to the paper fibers and $\perp$ the direction perpendicular (see Ref. [32] Matsumoto \textit{et al.}). Here, the paper sheets are cut such as the fibers are aligned with $L$ which means that the smaller modulus $E_\perp$ appears in the transverse bending energy term. Figure \ref{fig:state_diag_anis} represents the phase diagram obtained with $E_\perp/E_\parallel = 0.71$. This value is chosen to match the thin-sheet scaling regime with prefactor 3.7 found in the experiments.  

\begin{figure}[htbp]
    \centering
    \includegraphics[width=8.5cm]{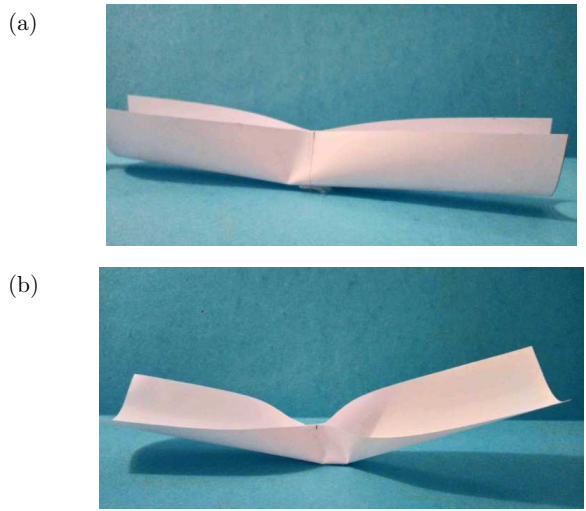}
    \caption{Morphology of a paper sheet with mid-length points contact for a sheet with (a) the paper fibers in the longitudinal direction and (b) the fibers in the transverse direction. An small adhesive strip is used to maintain the paper in self-contact.}
    \label{fig:s2}
\end{figure}

The anisotropy of paper has an impact on the buckling threshold and consistently on the post-buckling morphology. 
 Figure \ref{fig:s2} illustrates two morphologies obtained with a paper sheet of dimension $L=21$ cm, $W=6$ cm for the paper cut with (a) the fibers in the longitudinal direction (same orientation as in figure \ref{fig:s1}) and (b) the fibers in the transverse direction. From a visual comparison, the sheet (a) with the fibers parallel to $L$ is a higher state of crumpling than (b). This is confirmed by the number of d-cones (4 for (a) and 2 for (b)) and recess angle of the longitudinal centerline $v=0$.

\section{Spring network model}
Figure \ref{fig:s3} (a) represents a unit cell made of 8 vertices arranged in a cubic structure. Two types of linear springs are used:
\begin{itemize}
    \item springs of length 1 with a spring constant $k_1$ for the first neighbours (solid lines)
    \item springs of length $\sqrt{2}$ with a spring constant $k_2$ for the second-nearest neighbours (dashed lines)
\end{itemize}
\subsection{Structure}
\begin{figure}[htbp]
    \centering
    \includegraphics[width=8.5cm]{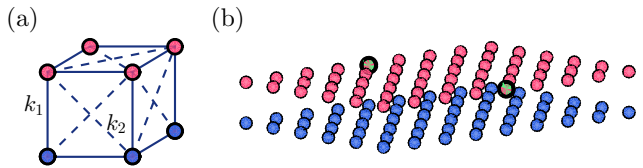}
    \caption{(a) Schematics of unit cell with $2\times 2 \times 2$ vertices. The two colors indicates the upper and lower layers. The first neighbours are connected by linear springs $k_1$ with length 1 and the second neighbours on faces are connected by springs $k_2$ of length $\sqrt{2}$. (b) Representation of a sheet with $9 \times 5 \times 2$. The mid-length vertices are highlighted in green.   }
    \label{fig:s3}
\end{figure}

Figure \ref{fig:s3} (b) represents a small sheet  made of $9 \times 5 \times 2$ vertices. The structure is at rest without additional constraints and the elastic energy is nil (all springs have a length 1 or $\sqrt{2}$). The vertices at mid-length are highlighted on the upper face. To force self-contact and bending, a supplementary spring linking the two mid-length points is added.

\subsection{Spring constants}
\begin{figure}[htbp]
    \centering
    \includegraphics[width=8.5cm]{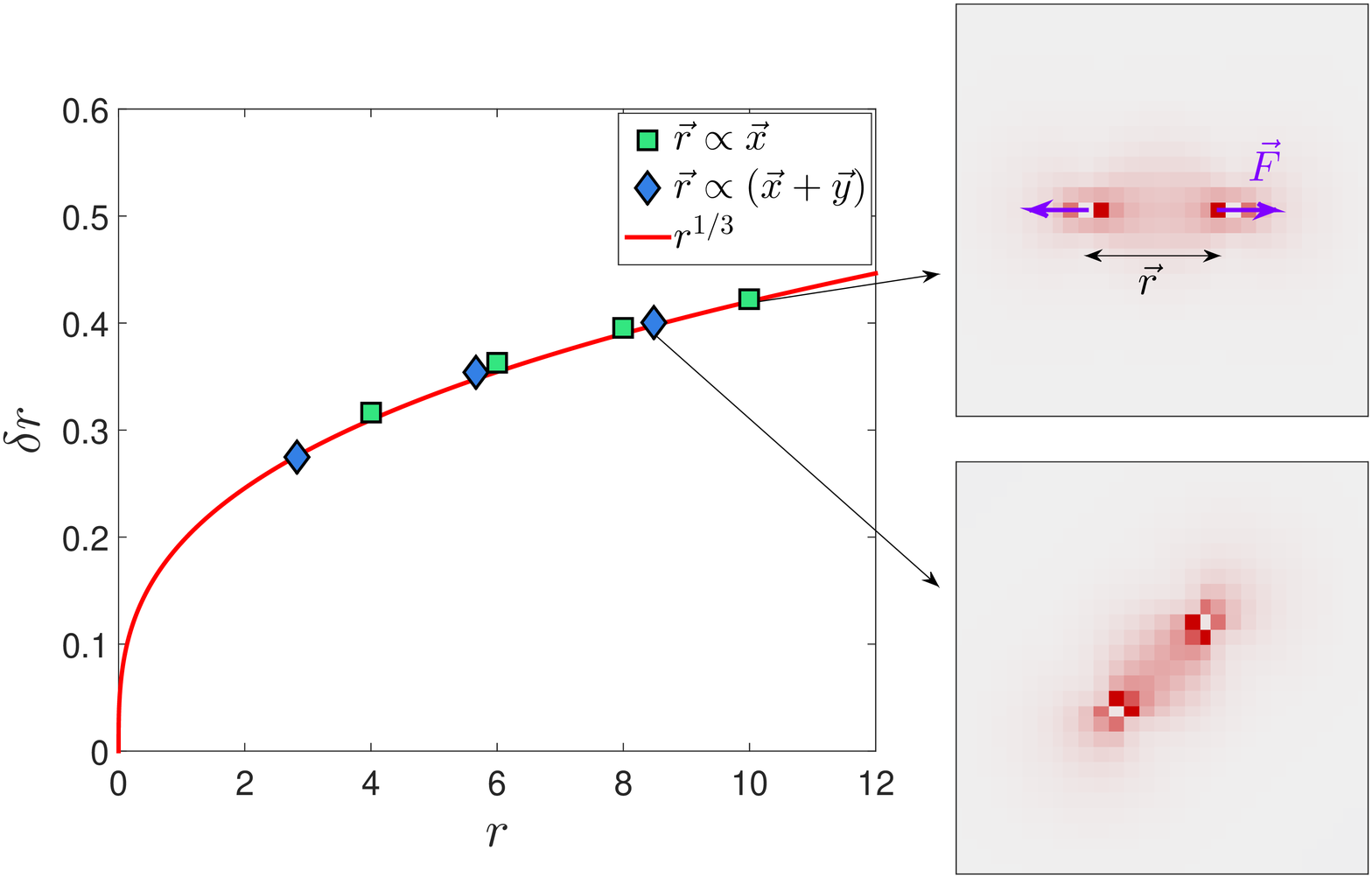}
    \caption{Displacement $\delta r$ of a pair of points separated by a distance $r$ for a stretching force of magnitude $F$. The size of the sheet is $81 \times 81 \times 2$ vertices and $r$ is small enough ($r\leq 10$) to avoid finite size effect of the sheet. The displacement is computed for the force $F$ either in a longitudinal direction (squares, $F$ parallel to springs $k_1$) or in a diagonal direction (diamonds, $F$ parallel to springs $k_2$). The inserted images represents the elastic energy for the $15 \times 15$ vertices in the middle of the sheet.}
    \label{fig:s4}
\end{figure}

The value of the springs constants $k_1$ and $k_2$ is set to have the same force-displacement relation for the stretching along the sides and the diagonals of the cubic cells. Figure \ref{fig:s4} presents the displacements of two points at a distance $r$ for an applied force $F$ and $-F$ in the cases where the two points are on the longitudinal direction $\vec{x}$ and on the diagonal direction $\vec{x}+\vec{y}$.
The superimposition of the simulation results for the force applied in the longitudinal and the diagonal directions is obtained for $k_1/k_2 = \sqrt{2}$.

\subsection{Mechanical thickness}
\begin{figure}[htbp]
    \centering
    \includegraphics[width=8.0cm]{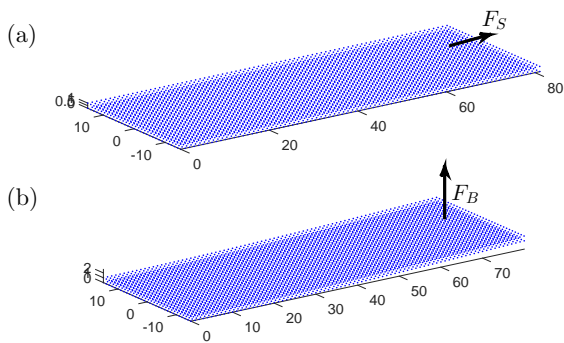}
    \caption{Sheet made of $81 \times 31 \times 2$ vertices submitted to (a) a stretching deformation and (b) a bending deformation. The forces are such as $F_S/F_B = L^2$  ($L = 80$) and they are distributed on the width of the strip (only one force vector is represented for clarity). The displacements for the extension (a) and the deflection (b) are used to calibrate the effective thickness of the sheet according to eq. \ref{eq:F_ratio2}  }
    \label{fig:s5}
\end{figure}

The thickness of the spring-network is calibrated by the simulation of stretching and bending tests. Figure \ref{fig:s5} represents a sheet made of $81 \times 31 \times 2$ vertices submitted to (a) an in-plane stretching force $F_S$ and (b) to an out-of-plane bending force $F_B$.  

For an elastic plate of dimension $L\times W \times t$, the stretching force-displacement relation is given by:
\begin{equation}
F_S = E\frac{Wt}{L}\delta_S\label{eq:F_S}
\end{equation}
with $\delta_S$ the displacement parallel to the applied force $F_S$ and $t$ the thickness.

The measurement of the displacement orthogonal to the stretching direction $\delta_S^*$ defines the Poisson ratio $\nu=\delta_S^*/\delta_S =0.29$.
The bending force-displacement relation reads
\begin{equation}
F_B = \frac{3EI}{L^3}\delta_B \quad \mathrm{with} \quad I = \frac{EW{t}^3}{12(1-\nu^2)}\, .
\label{eq:F_B}
\end{equation}

Combining eqs. (\ref{eq:F_S}) and (\ref{eq:F_B}), one can get the ratio
\begin{equation}
\frac{F_B}{F_S} = \frac{1}{4 (1-\nu^2)}\left(\frac{t}{L}\right)^2\frac{\delta_B}{\delta_S}\label{eq:F_ratio}
\end{equation}
One obtains the effective thickness of the two-layer sheet, 
\begin{equation}
t = \sqrt{4(1-\nu^2) L^2\frac{F_B}{F_S}\frac{\delta_S}{\delta_B}}\label{eq:F_ratio2}
\end{equation}

The simulation presented in figure \ref{fig:s5} is performed with ${F_B}/{F_S} = (1/L)^2$, with $L=80$ for the 80 intervals between the 81 vertices on the longitudinal direction. The amplitudes obtained in the simulation are $\delta_S =    1.0897$ and $\delta_B =  1.7224$.
According to eq. (\ref{eq:F_ratio2}), the thickness of the spring model is 
\begin{equation}
t = 1.52
\label{eq:t_e}
\end{equation}

\subsection{Dynamics of the vertices and equilibrium}\label{app:dyn}
The dynamics of the vertices is numerically solved by iteration of the forces on each vertex. The equation of motion for each vertex is:
\begin{equation}
  m\frac{d^2\vec{r}_i}{dt^2} = -\Gamma \frac{d\vec{r}_i}{dt} - \Sigma_{j} \vec{F_{ij}}\label{eq:mot}
\end{equation}
with $\vec{r}_i$ the location of the vertex $i$, $m=1$ the mass of the vertex, $\Gamma$ a damping parameter and $\Sigma_j$ the force from the springs $k_1$ and $k_2$ connecting the neighbours vertices $j$ to the vertex $i$. The damping coefficient $\Gamma$ is in the range $10^{-4}$ to $0.5$, depending on the network size. 
\begin{figure}[htbp]
    \centering
    \includegraphics[width=8.5cm]{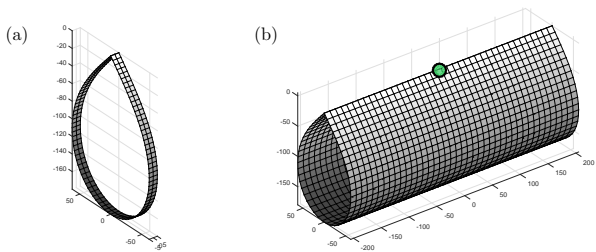}
    \caption{(a) Short-length sheet with $11 \times 471 \times 2$ vertices with mid-length points contact. (b) Sheet obtained by longitudinal replication of a transverse profile $1 \times 471 \times 2$ from (a). The sheet has a total $1701 \times 471 \times 2$ vertices.}
    \label{fig:s6}
\end{figure}

A simple first-order integration method is used to solve the equation of motion (\ref{eq:mot}). This method is known to produce inaccurate time resolution of differential equations. However, the main objective of this simulation is to identify the final state of the sheet. To minimize the number of iterative steps to converge towards equilibrium, a large integration step is taken with $\Delta t = 1$ for spring constants $k_1 = 0.3568$ and $k_2 = 0.2523$, satisfying the condition $k_1 = \sqrt{2}k_2$. A very small random noise is added to the position (of the order of $10^{-6}$ as compared to 1 the typical distance between two vertices) in order to avoid symmetry breaking degeneracy issues.

The sheet profiles are first computed for short-length sheets. Figure \ref{fig:s6} (a) presents the stationary configuration of a sheet made of $11 \times 471 \times 2$ vertices after multiple iteration of the equation of motion for each vertices ($2\:10^4$ iteration steps). In the limit of small length, the sheet configuration matches with Euler's elastica model for slender structures (pure bending limit). 
Figure \ref{fig:s6} (b) presents the initial state of the sheet's configuration with $1701 \times 471 \times 2$ obtained by replication of the transverse profile in figure \ref{fig:s6} (a).
The simulation of the equation of motion is performed by maintaining a contact between the two mid-length points marked by a circle in figure \ref{fig:s6} (b). 

\begin{figure}[htbp]
    \centering
    \includegraphics[width=8.5cm]{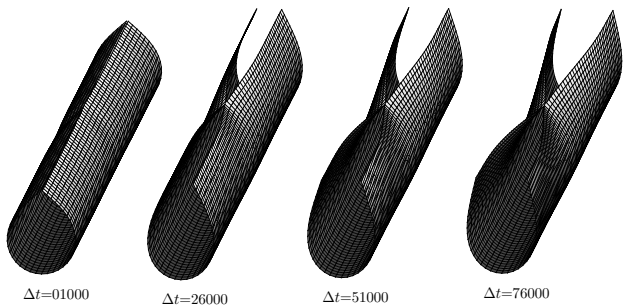}
    \caption{Sequence showing the evolution of a sheet of $1401 \times 401 \times 2$ vertices during the simulation of the spring network model. Time step $\Delta t/1000 = 1, 26, 51, 76$.}
    \label{fig:s7}
\end{figure}

Figure \ref{fig:s7} presents the configuration of a sheet of dimensions $1401 \times 401 \times 2$ vertices during the simulation after a number of iteration steps $\Delta t$.
In principle, the simulations should be run until equilibrium is reached. Because simulations are time consuming, they are stopped when buckling and stress focusing appear: the goal of the simulation is to identify the existence of buckling and not to follow in detail the post-buckling equilibrium shape. In figure 3 (c) of the main article, the simulation points for the smooth opening (circles) are obtained for sheets at mechanical equilibrium. For the data points corresponding to buckled sheets (squares and star), equilibrium is not reached for some of the largest sheets. It does not affect the validity of the threshold since the minimal stretching keeps on decreasing during stress focusing.

\subsection{Dynamics and threshold}
\begin{figure}[htbp]
    \centering
    \includegraphics[width=8.5cm]{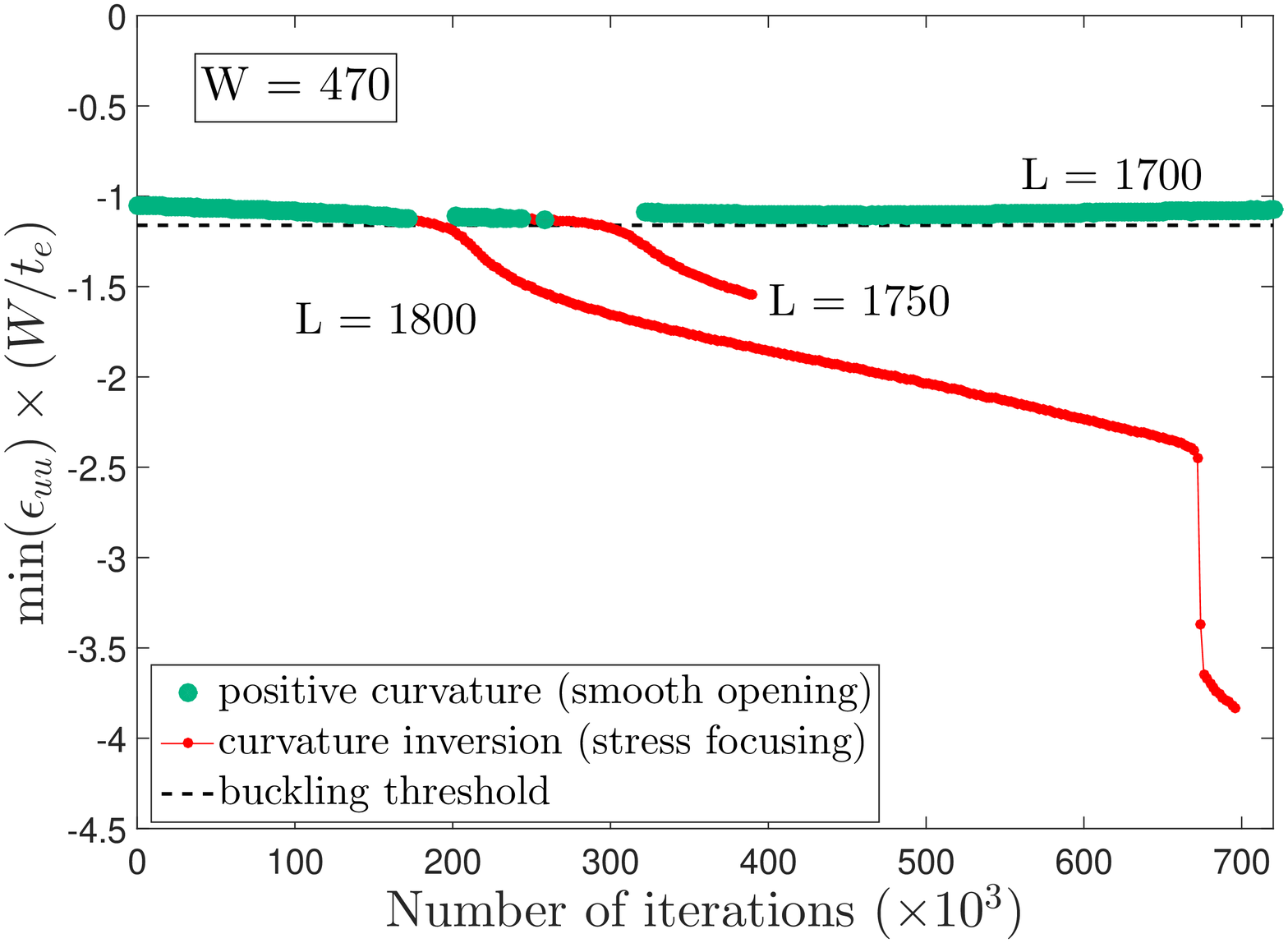}
    \caption{Evolution of the minimal strain (longitudinal compression) during 3 simulations with 471 vertices for the transverse direction and 1701, 1751 and 1800 vertices for the longitudinal direction. The dashed line is the threshold value -1.16. A video file \cite{video} shows the evolution of the sheet with $L=1800$. The abrupt transition corresponds to the transition between 4 d-cones and 2 d-cones.}
    \label{fig:s11_minstrain}
\end{figure}

Figure \ref{fig:s11_minstrain} presents the evolution of the minimal strain $\epsilon_{uu}\times (W/t)$ for 3 sheet sizes as a function of the simulation time step. The 3 sheets have 471 vertices in the transverse direction and 1701, 1751 and 1801 vertices in the longitudinal direction. The data for 1751 vertices is shifted horizontally for clarity. 

For the shortest sheet $1701 \times 471 \times 2$, the normalized strain stays above the threshold marked by a dashed line. For 1751 and 1801, the strain is decreasing with the iteration number (opening of the sheet represented in figure \ref{fig:s7}). The color and symbol change indicates a sign change for the curvature of the sheet transverse profiles at $u=\delta/2$ (see criteria in figure 3 (c) of the main article and figure 3 (b) for the definition of $\delta$).

The simulation with 1801 vertices explores a longer simulation time and a second and more abrupt transition is observed after $675\times 10^3$ time steps and the minimal strain almost double. This second transition is associated with the traveling and merging of the stress focusing points (figure 3 (d) with the separation $\delta$). The video \cite{video} shows the evolution of the sheet profile $1751 \times 471 \times 2$ and the two transitions (first transition around $300\times 10^3$ and second, more abrupt transition at $675\times 10^3$).
\begin{figure}[htbp]
    \centering
    \includegraphics[width=8.5cm]{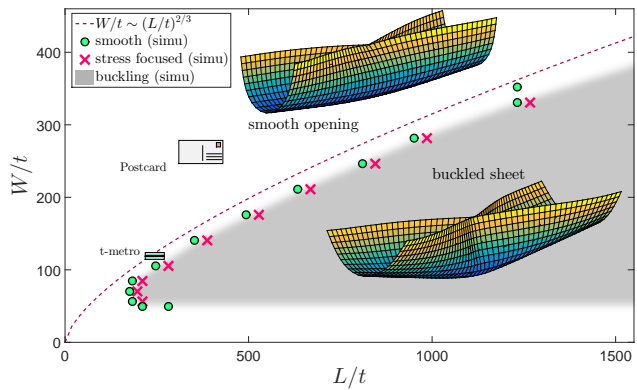}
    \caption{Representation of the simulation results to obtain the parameter region $(L/t, W/t)$ with stress focusing. The experimental points are not presented to avoid confusion. The dashed line is the same as in figure 2 of the main article. }
    \label{fig:s9}
\end{figure}

Although it is close, the curvature inversion does not exactly match with the threshold $-1.24$. One should expect that the compression threshold depends on the local curvature of the sheet. Since the location of the minimal strain is not self-similar (see figure 3 (d) of the main article with $\delta(W)$), one should expect that the appropriate normalization of $\epsilon_{uu}$ is not exactly $W/t$ but a slightly modified function that accounts for the parameter $\delta$, at the second order. Besides, the simulation includes added noise, which could explain fluctuating dynamics, notably if an instability is about to be reached (see the plot for 1851 vertices with the curvature sign that fluctuates around $50-100\times 10^3$ times steps).

The simulation $1701 \times 471 \times 2$ was performed from an initial configuration taken from the sheet $1751 \times 471 \times 2$ by removing 25 vertices on both ends, just before the transition. This is why the minimum strain increases for the simulation $1701 \times 471 \times 2$. This strategy was done to be sure that the smooth opening is stable for the configuration $1701 \times 471 \times 2$. 

\subsection{State diagram}
The simulation of the spring network model for different sheets sizes is used to establish the buckling region in parameters ($L/t$, $W/t$). Figure \ref{fig:s9} is the same figure that figure 3 in the main article excepted that the simulation data are presented. The experimental data for the acetate sheets are removed from the plot (those data points are distributed around the dashed line).

\bibliography{biblio}

\begin{thebibliography}{47}%
\makeatletter
\providecommand \@ifxundefined [1]{%
 \@ifx{#1\undefined}
}%
\providecommand \@ifnum [1]{%
 \ifnum #1\expandafter \@firstoftwo
 \else \expandafter \@secondoftwo
 \fi
}%
\providecommand \@ifx [1]{%
 \ifx #1\expandafter \@firstoftwo
 \else \expandafter \@secondoftwo
 \fi
}%
\providecommand \natexlab [1]{#1}%
\providecommand \enquote  [1]{``#1''}%
\providecommand \bibnamefont  [1]{#1}%
\providecommand \bibfnamefont [1]{#1}%
\providecommand \citenamefont [1]{#1}%
\providecommand \href@noop [0]{\@secondoftwo}%
\providecommand \href [0]{\begingroup \@sanitize@url \@href}%
\providecommand \@href[1]{\@@startlink{#1}\@@href}%
\providecommand \@@href[1]{\endgroup#1\@@endlink}%
\providecommand \@sanitize@url [0]{\catcode `\\12\catcode `\$12\catcode
  `\&12\catcode `\#12\catcode `\^12\catcode `\_12\catcode `\%12\relax}%
\providecommand \@@startlink[1]{}%
\providecommand \@@endlink[0]{}%
\providecommand \url  [0]{\begingroup\@sanitize@url \@url }%
\providecommand \@url [1]{\endgroup\@href {#1}{\urlprefix }}%
\providecommand \urlprefix  [0]{URL }%
\providecommand \Eprint [0]{\href }%
\providecommand \doibase [0]{http://dx.doi.org/}%
\providecommand \selectlanguage [0]{\@gobble}%
\providecommand \bibinfo  [0]{\@secondoftwo}%
\providecommand \bibfield  [0]{\@secondoftwo}%
\providecommand \translation [1]{[#1]}%
\providecommand \BibitemOpen [0]{}%
\providecommand \bibitemStop [0]{}%
\providecommand \bibitemNoStop [0]{.\EOS\space}%
\providecommand \EOS [0]{\spacefactor3000\relax}%
\providecommand \BibitemShut  [1]{\csname bibitem#1\endcsname}%
\let\auto@bib@innerbib\@empty
\bibitem [{\citenamefont {Rayleigh}(1888)}]{rayleigh1888bending}%
  \BibitemOpen
  \bibfield  {author} {\bibinfo {author} {\bibfnamefont {Lord}\ \bibnamefont
  {Rayleigh}},\ }\bibfield  {title} {\enquote {\bibinfo {title} {On the bending
  and vibration of thin elastic shells, especially of cylindrical form},}\
  }\href@noop {} {\bibfield  {journal} {\bibinfo  {journal} {Proceedings of the
  Royal Society of London}\ }\textbf {\bibinfo {volume} {45}},\ \bibinfo
  {pages} {105--123} (\bibinfo {year} {1888})}\BibitemShut {NoStop}%
\bibitem [{\citenamefont {Witten}(2007)}]{witten2007stress}%
  \BibitemOpen
  \bibfield  {author} {\bibinfo {author} {\bibfnamefont {Thomas~A}\
  \bibnamefont {Witten}},\ }\bibfield  {title} {\enquote {\bibinfo {title}
  {Stress focusing in elastic sheets},}\ }\href@noop {} {\bibfield  {journal}
  {\bibinfo  {journal} {Reviews of Modern Physics}\ }\textbf {\bibinfo {volume}
  {79}},\ \bibinfo {pages} {643} (\bibinfo {year} {2007})}\BibitemShut
  {NoStop}%
\bibitem [{\citenamefont {Cerda}\ and\ \citenamefont
  {Mahadevan}(1998)}]{cerda1998conical}%
  \BibitemOpen
  \bibfield  {author} {\bibinfo {author} {\bibfnamefont {Enrique}\ \bibnamefont
  {Cerda}}\ and\ \bibinfo {author} {\bibfnamefont {L}~\bibnamefont
  {Mahadevan}},\ }\bibfield  {title} {\enquote {\bibinfo {title} {Conical
  surfaces and crescent singularities in crumpled sheets},}\ }\href@noop {}
  {\bibfield  {journal} {\bibinfo  {journal} {Physical Review Letters}\
  }\textbf {\bibinfo {volume} {80}},\ \bibinfo {pages} {2358} (\bibinfo {year}
  {1998})}\BibitemShut {NoStop}%
\bibitem [{\citenamefont {Cha{\"\i}eb}\ \emph {et~al.}(1998)\citenamefont
  {Cha{\"\i}eb}, \citenamefont {Melo},\ and\ \citenamefont
  {G{\'e}minard}}]{chaieb1998experimental}%
  \BibitemOpen
  \bibfield  {author} {\bibinfo {author} {\bibfnamefont {Sahraoui}\
  \bibnamefont {Cha{\"\i}eb}}, \bibinfo {author} {\bibfnamefont {Francisco}\
  \bibnamefont {Melo}}, \ and\ \bibinfo {author} {\bibfnamefont
  {Jean-Christophe}\ \bibnamefont {G{\'e}minard}},\ }\bibfield  {title}
  {\enquote {\bibinfo {title} {Experimental study of developable cones},}\
  }\href@noop {} {\bibfield  {journal} {\bibinfo  {journal} {Physical review
  letters}\ }\textbf {\bibinfo {volume} {80}},\ \bibinfo {pages} {2354}
  (\bibinfo {year} {1998})}\BibitemShut {NoStop}%
\bibitem [{\citenamefont {Cerda}\ \emph {et~al.}(1999)\citenamefont {Cerda},
  \citenamefont {Chaieb}, \citenamefont {Melo},\ and\ \citenamefont
  {Mahadevan}}]{cerda1999conical}%
  \BibitemOpen
  \bibfield  {author} {\bibinfo {author} {\bibfnamefont {Enrique}\ \bibnamefont
  {Cerda}}, \bibinfo {author} {\bibfnamefont {Sahraoui}\ \bibnamefont
  {Chaieb}}, \bibinfo {author} {\bibfnamefont {Francisco}\ \bibnamefont
  {Melo}}, \ and\ \bibinfo {author} {\bibfnamefont {L}~\bibnamefont
  {Mahadevan}},\ }\bibfield  {title} {\enquote {\bibinfo {title} {Conical
  dislocations in crumpling},}\ }\href@noop {} {\bibfield  {journal} {\bibinfo
  {journal} {Nature}\ }\textbf {\bibinfo {volume} {401}},\ \bibinfo {pages}
  {46--49} (\bibinfo {year} {1999})}\BibitemShut {NoStop}%
\bibitem [{\citenamefont {Ben~Amar}\ and\ \citenamefont
  {Pomeau}(1997)}]{ben1997crumpled}%
  \BibitemOpen
  \bibfield  {author} {\bibinfo {author} {\bibfnamefont {M}~\bibnamefont
  {Ben~Amar}}\ and\ \bibinfo {author} {\bibfnamefont {Y}~\bibnamefont
  {Pomeau}},\ }\bibfield  {title} {\enquote {\bibinfo {title} {Crumpled
  paper},}\ }\href@noop {} {\bibfield  {journal} {\bibinfo  {journal}
  {Proceedings of the Royal Society of London. Series A: Mathematical, Physical
  and Engineering Sciences}\ }\textbf {\bibinfo {volume} {453}},\ \bibinfo
  {pages} {729--755} (\bibinfo {year} {1997})}\BibitemShut {NoStop}%
\bibitem [{\citenamefont {Lobkovsky}\ \emph {et~al.}(1995)\citenamefont
  {Lobkovsky}, \citenamefont {Gentges}, \citenamefont {Li}, \citenamefont
  {Morse},\ and\ \citenamefont {Witten}}]{lobkovsky1995scaling}%
  \BibitemOpen
  \bibfield  {author} {\bibinfo {author} {\bibfnamefont {Alex}\ \bibnamefont
  {Lobkovsky}}, \bibinfo {author} {\bibfnamefont {Sharon}\ \bibnamefont
  {Gentges}}, \bibinfo {author} {\bibfnamefont {Hao}\ \bibnamefont {Li}},
  \bibinfo {author} {\bibfnamefont {David}\ \bibnamefont {Morse}}, \ and\
  \bibinfo {author} {\bibfnamefont {Thomas~A}\ \bibnamefont {Witten}},\
  }\bibfield  {title} {\enquote {\bibinfo {title} {Scaling properties of
  stretching ridges in a crumpled elastic sheet},}\ }\href@noop {} {\bibfield
  {journal} {\bibinfo  {journal} {Science}\ }\textbf {\bibinfo {volume}
  {270}},\ \bibinfo {pages} {1482--1485} (\bibinfo {year} {1995})}\BibitemShut
  {NoStop}%
\bibitem [{\citenamefont {Sultan}\ and\ \citenamefont
  {Boudaoud}(2006)}]{sultan2006statistics}%
  \BibitemOpen
  \bibfield  {author} {\bibinfo {author} {\bibfnamefont {Eric}\ \bibnamefont
  {Sultan}}\ and\ \bibinfo {author} {\bibfnamefont {Arezki}\ \bibnamefont
  {Boudaoud}},\ }\bibfield  {title} {\enquote {\bibinfo {title} {Statistics of
  crumpled paper},}\ }\href@noop {} {\bibfield  {journal} {\bibinfo  {journal}
  {Physical review letters}\ }\textbf {\bibinfo {volume} {96}},\ \bibinfo
  {pages} {136103} (\bibinfo {year} {2006})}\BibitemShut {NoStop}%
\bibitem [{\citenamefont {Andresen}\ \emph {et~al.}(2007)\citenamefont
  {Andresen}, \citenamefont {Hansen},\ and\ \citenamefont
  {Schmittbuhl}}]{andresen2007ridge}%
  \BibitemOpen
  \bibfield  {author} {\bibinfo {author} {\bibfnamefont {Christian~Andr{\'e}}\
  \bibnamefont {Andresen}}, \bibinfo {author} {\bibfnamefont {Alex}\
  \bibnamefont {Hansen}}, \ and\ \bibinfo {author} {\bibfnamefont {Jean}\
  \bibnamefont {Schmittbuhl}},\ }\bibfield  {title} {\enquote {\bibinfo {title}
  {Ridge network in crumpled paper},}\ }\href@noop {} {\bibfield  {journal}
  {\bibinfo  {journal} {Physical review E}\ }\textbf {\bibinfo {volume} {76}},\
  \bibinfo {pages} {026108} (\bibinfo {year} {2007})}\BibitemShut {NoStop}%
\bibitem [{\citenamefont {Balankin}\ \emph {et~al.}(2010)\citenamefont
  {Balankin}, \citenamefont {Ochoa}, \citenamefont {Miguel}, \citenamefont
  {Ortiz},\ and\ \citenamefont {Cruz}}]{balankin2010fractal}%
  \BibitemOpen
  \bibfield  {author} {\bibinfo {author} {\bibfnamefont {Alexander~S}\
  \bibnamefont {Balankin}}, \bibinfo {author} {\bibfnamefont {Didier~Samayoa}\
  \bibnamefont {Ochoa}}, \bibinfo {author} {\bibfnamefont {Israel~Andr{\'e}s}\
  \bibnamefont {Miguel}}, \bibinfo {author} {\bibfnamefont
  {Juli{\'a}n~Pati{\~n}o}\ \bibnamefont {Ortiz}}, \ and\ \bibinfo {author}
  {\bibfnamefont {Miguel {\'A}ngel~Mart{\'\i}nez}\ \bibnamefont {Cruz}},\
  }\bibfield  {title} {\enquote {\bibinfo {title} {Fractal topology of
  hand-crumpled paper},}\ }\href@noop {} {\bibfield  {journal} {\bibinfo
  {journal} {Physical Review E}\ }\textbf {\bibinfo {volume} {81}},\ \bibinfo
  {pages} {061126} (\bibinfo {year} {2010})}\BibitemShut {NoStop}%
\bibitem [{\citenamefont {Cambou}\ and\ \citenamefont
  {Menon}(2011)}]{cambou2011three}%
  \BibitemOpen
  \bibfield  {author} {\bibinfo {author} {\bibfnamefont {Anne~Dominique}\
  \bibnamefont {Cambou}}\ and\ \bibinfo {author} {\bibfnamefont {Narayanan}\
  \bibnamefont {Menon}},\ }\bibfield  {title} {\enquote {\bibinfo {title}
  {Three-dimensional structure of a sheet crumpled into a ball},}\ }\href@noop
  {} {\bibfield  {journal} {\bibinfo  {journal} {Proceedings of the National
  Academy of Sciences}\ }\textbf {\bibinfo {volume} {108}},\ \bibinfo {pages}
  {14741--14745} (\bibinfo {year} {2011})}\BibitemShut {NoStop}%
\bibitem [{\citenamefont {Croll}\ \emph {et~al.}(2019)\citenamefont {Croll},
  \citenamefont {Twohig},\ and\ \citenamefont {Elder}}]{croll2019compressive}%
  \BibitemOpen
  \bibfield  {author} {\bibinfo {author} {\bibfnamefont {Andrew~B}\
  \bibnamefont {Croll}}, \bibinfo {author} {\bibfnamefont {Timothy}\
  \bibnamefont {Twohig}}, \ and\ \bibinfo {author} {\bibfnamefont {Theresa}\
  \bibnamefont {Elder}},\ }\bibfield  {title} {\enquote {\bibinfo {title} {The
  compressive strength of crumpled matter},}\ }\href@noop {} {\bibfield
  {journal} {\bibinfo  {journal} {Nature communications}\ }\textbf {\bibinfo
  {volume} {10}},\ \bibinfo {pages} {1--8} (\bibinfo {year}
  {2019})}\BibitemShut {NoStop}%
\bibitem [{\citenamefont {Gottesman}\ \emph {et~al.}(2015)\citenamefont
  {Gottesman}, \citenamefont {Efrati},\ and\ \citenamefont
  {Rubinstein}}]{gottesman2015furrows}%
  \BibitemOpen
  \bibfield  {author} {\bibinfo {author} {\bibfnamefont {Omer}\ \bibnamefont
  {Gottesman}}, \bibinfo {author} {\bibfnamefont {Efi}\ \bibnamefont {Efrati}},
  \ and\ \bibinfo {author} {\bibfnamefont {Shmuel~M}\ \bibnamefont
  {Rubinstein}},\ }\bibfield  {title} {\enquote {\bibinfo {title} {Furrows in
  the wake of propagating d-cones},}\ }\href@noop {} {\bibfield  {journal}
  {\bibinfo  {journal} {Nature communications}\ }\textbf {\bibinfo {volume}
  {6}},\ \bibinfo {pages} {1--7} (\bibinfo {year} {2015})}\BibitemShut
  {NoStop}%
\bibitem [{\citenamefont {Chopin}\ and\ \citenamefont
  {Kudrolli}(2016)}]{chopin2016disclinations}%
  \BibitemOpen
  \bibfield  {author} {\bibinfo {author} {\bibfnamefont {Julien}\ \bibnamefont
  {Chopin}}\ and\ \bibinfo {author} {\bibfnamefont {Arshad}\ \bibnamefont
  {Kudrolli}},\ }\bibfield  {title} {\enquote {\bibinfo {title} {Disclinations,
  e-cones, and their interactions in extensible sheets},}\ }\href@noop {}
  {\bibfield  {journal} {\bibinfo  {journal} {Soft Matter}\ }\textbf {\bibinfo
  {volume} {12}},\ \bibinfo {pages} {4457--4462} (\bibinfo {year}
  {2016})}\BibitemShut {NoStop}%
\bibitem [{\citenamefont {Rogers}\ and\ \citenamefont
  {Huang}(2009)}]{rogers2009curvy}%
  \BibitemOpen
  \bibfield  {author} {\bibinfo {author} {\bibfnamefont {John~A}\ \bibnamefont
  {Rogers}}\ and\ \bibinfo {author} {\bibfnamefont {Yonggang}\ \bibnamefont
  {Huang}},\ }\bibfield  {title} {\enquote {\bibinfo {title} {A curvy, stretchy
  future for electronics},}\ }\href@noop {} {\bibfield  {journal} {\bibinfo
  {journal} {Proceedings of the National Academy of Sciences}\ }\textbf
  {\bibinfo {volume} {106}},\ \bibinfo {pages} {10875--10876} (\bibinfo {year}
  {2009})}\BibitemShut {NoStop}%
\bibitem [{\citenamefont {Luo}\ \emph {et~al.}(2011)\citenamefont {Luo},
  \citenamefont {Jang}, \citenamefont {Sun}, \citenamefont {Xiao},
  \citenamefont {He}, \citenamefont {Katsoulidis}, \citenamefont {Kanatzidis},
  \citenamefont {Gibson},\ and\ \citenamefont {Huang}}]{luo2011compression}%
  \BibitemOpen
  \bibfield  {author} {\bibinfo {author} {\bibfnamefont {Jiayan}\ \bibnamefont
  {Luo}}, \bibinfo {author} {\bibfnamefont {Hee~Dong}\ \bibnamefont {Jang}},
  \bibinfo {author} {\bibfnamefont {Tao}\ \bibnamefont {Sun}}, \bibinfo
  {author} {\bibfnamefont {Li}~\bibnamefont {Xiao}}, \bibinfo {author}
  {\bibfnamefont {Zhen}\ \bibnamefont {He}}, \bibinfo {author} {\bibfnamefont
  {Alexandros~P}\ \bibnamefont {Katsoulidis}}, \bibinfo {author} {\bibfnamefont
  {Mercouri~G}\ \bibnamefont {Kanatzidis}}, \bibinfo {author} {\bibfnamefont
  {J~Murray}\ \bibnamefont {Gibson}}, \ and\ \bibinfo {author} {\bibfnamefont
  {Jiaxing}\ \bibnamefont {Huang}},\ }\bibfield  {title} {\enquote {\bibinfo
  {title} {Compression and aggregation-resistant particles of crumpled soft
  sheets},}\ }\href@noop {} {\bibfield  {journal} {\bibinfo  {journal} {ACS
  nano}\ }\textbf {\bibinfo {volume} {5}},\ \bibinfo {pages} {8943--8949}
  (\bibinfo {year} {2011})}\BibitemShut {NoStop}%
\bibitem [{\citenamefont {Mao}\ \emph {et~al.}(2012)\citenamefont {Mao},
  \citenamefont {Wen}, \citenamefont {Kim}, \citenamefont {Lu}, \citenamefont
  {Hurley},\ and\ \citenamefont {Chen}}]{mao2012general}%
  \BibitemOpen
  \bibfield  {author} {\bibinfo {author} {\bibfnamefont {Shun}\ \bibnamefont
  {Mao}}, \bibinfo {author} {\bibfnamefont {Zhenhai}\ \bibnamefont {Wen}},
  \bibinfo {author} {\bibfnamefont {Haejune}\ \bibnamefont {Kim}}, \bibinfo
  {author} {\bibfnamefont {Ganhua}\ \bibnamefont {Lu}}, \bibinfo {author}
  {\bibfnamefont {Patrick}\ \bibnamefont {Hurley}}, \ and\ \bibinfo {author}
  {\bibfnamefont {Junhong}\ \bibnamefont {Chen}},\ }\bibfield  {title}
  {\enquote {\bibinfo {title} {A general approach to one-pot fabrication of
  crumpled graphene-based nanohybrids for energy applications},}\ }\href@noop
  {} {\bibfield  {journal} {\bibinfo  {journal} {ACS nano}\ }\textbf {\bibinfo
  {volume} {6}},\ \bibinfo {pages} {7505--7513} (\bibinfo {year}
  {2012})}\BibitemShut {NoStop}%
\bibitem [{\citenamefont {Ma}\ \emph {et~al.}(2012)\citenamefont {Ma},
  \citenamefont {Zachariah},\ and\ \citenamefont
  {Zangmeister}}]{ma2012crumpled}%
  \BibitemOpen
  \bibfield  {author} {\bibinfo {author} {\bibfnamefont {Xiaofei}\ \bibnamefont
  {Ma}}, \bibinfo {author} {\bibfnamefont {Michael~R}\ \bibnamefont
  {Zachariah}}, \ and\ \bibinfo {author} {\bibfnamefont {Christopher~D}\
  \bibnamefont {Zangmeister}},\ }\bibfield  {title} {\enquote {\bibinfo {title}
  {Crumpled nanopaper from graphene oxide},}\ }\href@noop {} {\bibfield
  {journal} {\bibinfo  {journal} {Nano letters}\ }\textbf {\bibinfo {volume}
  {12}},\ \bibinfo {pages} {486--489} (\bibinfo {year} {2012})}\BibitemShut
  {NoStop}%
\bibitem [{\citenamefont {Yan}\ \emph {et~al.}(2013)\citenamefont {Yan},
  \citenamefont {He}, \citenamefont {Chu}, \citenamefont {Liu}, \citenamefont
  {Meng}, \citenamefont {Dou}, \citenamefont {Zhang}, \citenamefont {Liu},
  \citenamefont {Nie},\ and\ \citenamefont {He}}]{yan2013strain}%
  \BibitemOpen
  \bibfield  {author} {\bibinfo {author} {\bibfnamefont {Wei}\ \bibnamefont
  {Yan}}, \bibinfo {author} {\bibfnamefont {Wen-Yu}\ \bibnamefont {He}},
  \bibinfo {author} {\bibfnamefont {Zhao-Dong}\ \bibnamefont {Chu}}, \bibinfo
  {author} {\bibfnamefont {Mengxi}\ \bibnamefont {Liu}}, \bibinfo {author}
  {\bibfnamefont {Lan}\ \bibnamefont {Meng}}, \bibinfo {author} {\bibfnamefont
  {Rui-Fen}\ \bibnamefont {Dou}}, \bibinfo {author} {\bibfnamefont {Yanfeng}\
  \bibnamefont {Zhang}}, \bibinfo {author} {\bibfnamefont {Zhongfan}\
  \bibnamefont {Liu}}, \bibinfo {author} {\bibfnamefont {Jia-Cai}\ \bibnamefont
  {Nie}}, \ and\ \bibinfo {author} {\bibfnamefont {Lin}\ \bibnamefont {He}},\
  }\bibfield  {title} {\enquote {\bibinfo {title} {Strain and curvature induced
  evolution of electronic band structures in twisted graphene bilayer},}\
  }\href@noop {} {\bibfield  {journal} {\bibinfo  {journal} {Nature
  communications}\ }\textbf {\bibinfo {volume} {4}},\ \bibinfo {pages} {1--7}
  (\bibinfo {year} {2013})}\BibitemShut {NoStop}%
\bibitem [{\citenamefont {Reis}(2015)}]{reis2015perspective}%
  \BibitemOpen
  \bibfield  {author} {\bibinfo {author} {\bibfnamefont {Pedro~M}\ \bibnamefont
  {Reis}},\ }\bibfield  {title} {\enquote {\bibinfo {title} {A perspective on
  the revival of structural (in) stability with novel opportunities for
  function: from buckliphobia to buckliphilia},}\ }\href@noop {} {\bibfield
  {journal} {\bibinfo  {journal} {Journal of Applied Mechanics}\ }\textbf
  {\bibinfo {volume} {82}} (\bibinfo {year} {2015})}\BibitemShut {NoStop}%
\bibitem [{\citenamefont {Holmes}(2019)}]{holmes2019elasticity}%
  \BibitemOpen
  \bibfield  {author} {\bibinfo {author} {\bibfnamefont {Douglas~P}\
  \bibnamefont {Holmes}},\ }\bibfield  {title} {\enquote {\bibinfo {title}
  {Elasticity and stability of shape changing structures},}\ }\href@noop {}
  {\bibfield  {journal} {\bibinfo  {journal} {Current opinion in colloid \&
  interface science}\ } (\bibinfo {year} {2019})}\BibitemShut {NoStop}%
\bibitem [{\citenamefont {Schroll}\ \emph {et~al.}(2011)\citenamefont
  {Schroll}, \citenamefont {Katifori},\ and\ \citenamefont
  {Davidovitch}}]{schroll2011elastic}%
  \BibitemOpen
  \bibfield  {author} {\bibinfo {author} {\bibfnamefont {Robert~D}\
  \bibnamefont {Schroll}}, \bibinfo {author} {\bibfnamefont {Eleni}\
  \bibnamefont {Katifori}}, \ and\ \bibinfo {author} {\bibfnamefont {Benny}\
  \bibnamefont {Davidovitch}},\ }\bibfield  {title} {\enquote {\bibinfo {title}
  {Elastic building blocks for confined sheets},}\ }\href@noop {} {\bibfield
  {journal} {\bibinfo  {journal} {Physical review letters}\ }\textbf {\bibinfo
  {volume} {106}},\ \bibinfo {pages} {074301} (\bibinfo {year}
  {2011})}\BibitemShut {NoStop}%
\bibitem [{\citenamefont {Fuentealba}\ \emph {et~al.}(2015)\citenamefont
  {Fuentealba}, \citenamefont {Albarr{\'a}n}, \citenamefont {Hamm},\ and\
  \citenamefont {Cerda}}]{fuentealba2015transition}%
  \BibitemOpen
  \bibfield  {author} {\bibinfo {author} {\bibfnamefont {JF}~\bibnamefont
  {Fuentealba}}, \bibinfo {author} {\bibfnamefont {O}~\bibnamefont
  {Albarr{\'a}n}}, \bibinfo {author} {\bibfnamefont {E}~\bibnamefont {Hamm}}, \
  and\ \bibinfo {author} {\bibfnamefont {E}~\bibnamefont {Cerda}},\ }\bibfield
  {title} {\enquote {\bibinfo {title} {Transition from isometric to stretching
  ridges in thin elastic films},}\ }\href@noop {} {\bibfield  {journal}
  {\bibinfo  {journal} {Physical Review E}\ }\textbf {\bibinfo {volume} {91}},\
  \bibinfo {pages} {032407} (\bibinfo {year} {2015})}\BibitemShut {NoStop}%
\bibitem [{\citenamefont {Boudaoud}\ \emph {et~al.}(2000)\citenamefont
  {Boudaoud}, \citenamefont {Patr{\'\i}cio}, \citenamefont {Couder},\ and\
  \citenamefont {Amar}}]{boudaoud2000dynamics}%
  \BibitemOpen
  \bibfield  {author} {\bibinfo {author} {\bibfnamefont {Arezki}\ \bibnamefont
  {Boudaoud}}, \bibinfo {author} {\bibfnamefont {Pedro}\ \bibnamefont
  {Patr{\'\i}cio}}, \bibinfo {author} {\bibfnamefont {Yves}\ \bibnamefont
  {Couder}}, \ and\ \bibinfo {author} {\bibfnamefont {Martine~Ben}\
  \bibnamefont {Amar}},\ }\bibfield  {title} {\enquote {\bibinfo {title}
  {Dynamics of singularities in a constrained elastic plate},}\ }\href@noop {}
  {\bibfield  {journal} {\bibinfo  {journal} {Nature}\ }\textbf {\bibinfo
  {volume} {407}},\ \bibinfo {pages} {718--720} (\bibinfo {year}
  {2000})}\BibitemShut {NoStop}%
\bibitem [{\citenamefont {Roman}\ and\ \citenamefont
  {Pocheau}(2012)}]{roman2012stress}%
  \BibitemOpen
  \bibfield  {author} {\bibinfo {author} {\bibfnamefont {Beno{\^\i}t}\
  \bibnamefont {Roman}}\ and\ \bibinfo {author} {\bibfnamefont {Alain}\
  \bibnamefont {Pocheau}},\ }\bibfield  {title} {\enquote {\bibinfo {title}
  {Stress defocusing in anisotropic compaction of thin sheets},}\ }\href@noop
  {} {\bibfield  {journal} {\bibinfo  {journal} {Physical review letters}\
  }\textbf {\bibinfo {volume} {108}},\ \bibinfo {pages} {074301} (\bibinfo
  {year} {2012})}\BibitemShut {NoStop}%
\bibitem [{\citenamefont {Cerda}\ \emph {et~al.}(2004)\citenamefont {Cerda},
  \citenamefont {Mahadevan},\ and\ \citenamefont {Pasini}}]{cerda2004elements}%
  \BibitemOpen
  \bibfield  {author} {\bibinfo {author} {\bibfnamefont {Enrique}\ \bibnamefont
  {Cerda}}, \bibinfo {author} {\bibfnamefont {Lakshminarayanan}\ \bibnamefont
  {Mahadevan}}, \ and\ \bibinfo {author} {\bibfnamefont {Jos{\'e}~Miguel}\
  \bibnamefont {Pasini}},\ }\bibfield  {title} {\enquote {\bibinfo {title} {The
  elements of draping},}\ }\href@noop {} {\bibfield  {journal} {\bibinfo
  {journal} {Proceedings of the National Academy of Sciences}\ }\textbf
  {\bibinfo {volume} {101}},\ \bibinfo {pages} {1806--1810} (\bibinfo {year}
  {2004})}\BibitemShut {NoStop}%
\bibitem [{\citenamefont {Lobkovsky}\ and\ \citenamefont
  {Witten}(1997)}]{lobkovsky1997properties}%
  \BibitemOpen
  \bibfield  {author} {\bibinfo {author} {\bibfnamefont {Alexander~E}\
  \bibnamefont {Lobkovsky}}\ and\ \bibinfo {author} {\bibfnamefont
  {TA}~\bibnamefont {Witten}},\ }\bibfield  {title} {\enquote {\bibinfo {title}
  {Properties of ridges in elastic membranes},}\ }\href@noop {} {\bibfield
  {journal} {\bibinfo  {journal} {Physical Review E}\ }\textbf {\bibinfo
  {volume} {55}},\ \bibinfo {pages} {1577} (\bibinfo {year}
  {1997})}\BibitemShut {NoStop}%
\bibitem [{\citenamefont {Cerda}\ and\ \citenamefont
  {Mahadevan}(2003)}]{cerda2003geometry}%
  \BibitemOpen
  \bibfield  {author} {\bibinfo {author} {\bibfnamefont {Enrique}\ \bibnamefont
  {Cerda}}\ and\ \bibinfo {author} {\bibfnamefont {Lakshminarayanan}\
  \bibnamefont {Mahadevan}},\ }\bibfield  {title} {\enquote {\bibinfo {title}
  {Geometry and physics of wrinkling},}\ }\href@noop {} {\bibfield  {journal}
  {\bibinfo  {journal} {Physical review letters}\ }\textbf {\bibinfo {volume}
  {90}},\ \bibinfo {pages} {074302} (\bibinfo {year} {2003})}\BibitemShut
  {NoStop}%
\bibitem [{\citenamefont {Vandeparre}\ \emph {et~al.}(2011)\citenamefont
  {Vandeparre}, \citenamefont {Pi{\~n}eirua}, \citenamefont {Brau},
  \citenamefont {Roman}, \citenamefont {Bico}, \citenamefont {Gay},
  \citenamefont {Bao}, \citenamefont {Lau}, \citenamefont {Reis},\ and\
  \citenamefont {Damman}}]{vandeparre2011wrinkling}%
  \BibitemOpen
  \bibfield  {author} {\bibinfo {author} {\bibfnamefont {Hugues}\ \bibnamefont
  {Vandeparre}}, \bibinfo {author} {\bibfnamefont {Miguel}\ \bibnamefont
  {Pi{\~n}eirua}}, \bibinfo {author} {\bibfnamefont {Fabian}\ \bibnamefont
  {Brau}}, \bibinfo {author} {\bibfnamefont {Benoit}\ \bibnamefont {Roman}},
  \bibinfo {author} {\bibfnamefont {Jos{\'e}}\ \bibnamefont {Bico}}, \bibinfo
  {author} {\bibfnamefont {Cyprien}\ \bibnamefont {Gay}}, \bibinfo {author}
  {\bibfnamefont {Wenzhong}\ \bibnamefont {Bao}}, \bibinfo {author}
  {\bibfnamefont {Chun~Ning}\ \bibnamefont {Lau}}, \bibinfo {author}
  {\bibfnamefont {Pedro~M}\ \bibnamefont {Reis}}, \ and\ \bibinfo {author}
  {\bibfnamefont {Pascal}\ \bibnamefont {Damman}},\ }\bibfield  {title}
  {\enquote {\bibinfo {title} {Wrinkling hierarchy in constrained thin sheets
  from suspended graphene to curtains},}\ }\href@noop {} {\bibfield  {journal}
  {\bibinfo  {journal} {Physical Review Letters}\ }\textbf {\bibinfo {volume}
  {106}},\ \bibinfo {pages} {224301} (\bibinfo {year} {2011})}\BibitemShut
  {NoStop}%
\bibitem [{\citenamefont {Barois}\ \emph {et~al.}(2014)\citenamefont {Barois},
  \citenamefont {Tadrist}, \citenamefont {Quilliet},\ and\ \citenamefont
  {Forterre}}]{barois2014curved}%
  \BibitemOpen
  \bibfield  {author} {\bibinfo {author} {\bibfnamefont {Thomas}\ \bibnamefont
  {Barois}}, \bibinfo {author} {\bibfnamefont {Lo{\"\i}c}\ \bibnamefont
  {Tadrist}}, \bibinfo {author} {\bibfnamefont {Catherine}\ \bibnamefont
  {Quilliet}}, \ and\ \bibinfo {author} {\bibfnamefont {Yo{\"e}l}\ \bibnamefont
  {Forterre}},\ }\bibfield  {title} {\enquote {\bibinfo {title} {How a curved
  elastic strip opens},}\ }\href@noop {} {\bibfield  {journal} {\bibinfo
  {journal} {Physical review letters}\ }\textbf {\bibinfo {volume} {113}},\
  \bibinfo {pages} {214301} (\bibinfo {year} {2014})}\BibitemShut {NoStop}%
\bibitem [{\citenamefont {Pini}\ \emph {et~al.}(2016)\citenamefont {Pini},
  \citenamefont {Ruz}, \citenamefont {Kosaka}, \citenamefont {Malvar},
  \citenamefont {Calleja},\ and\ \citenamefont {Tamayo}}]{pini2016two}%
  \BibitemOpen
  \bibfield  {author} {\bibinfo {author} {\bibfnamefont {Valerio}\ \bibnamefont
  {Pini}}, \bibinfo {author} {\bibfnamefont {JJ}~\bibnamefont {Ruz}}, \bibinfo
  {author} {\bibfnamefont {Priscila~M}\ \bibnamefont {Kosaka}}, \bibinfo
  {author} {\bibfnamefont {O}~\bibnamefont {Malvar}}, \bibinfo {author}
  {\bibfnamefont {Montserrat}\ \bibnamefont {Calleja}}, \ and\ \bibinfo
  {author} {\bibfnamefont {J}~\bibnamefont {Tamayo}},\ }\bibfield  {title}
  {\enquote {\bibinfo {title} {How two-dimensional bending can extraordinarily
  stiffen thin sheets},}\ }\href@noop {} {\bibfield  {journal} {\bibinfo
  {journal} {Scientific reports}\ }\textbf {\bibinfo {volume} {6}},\ \bibinfo
  {pages} {1--6} (\bibinfo {year} {2016})}\BibitemShut {NoStop}%
\bibitem [{\citenamefont {Matsumoto}\ \emph {et~al.}(2018)\citenamefont
  {Matsumoto}, \citenamefont {Sano},\ and\ \citenamefont
  {Wada}}]{matsumoto2018pinching}%
  \BibitemOpen
  \bibfield  {author} {\bibinfo {author} {\bibfnamefont {Daichi}\ \bibnamefont
  {Matsumoto}}, \bibinfo {author} {\bibfnamefont {Tomohiko~G}\ \bibnamefont
  {Sano}}, \ and\ \bibinfo {author} {\bibfnamefont {Hirofumi}\ \bibnamefont
  {Wada}},\ }\bibfield  {title} {\enquote {\bibinfo {title} {Pinching an open
  cylindrical shell: Extended deformation and its persistence},}\ }\href@noop
  {} {\bibfield  {journal} {\bibinfo  {journal} {EPL (Europhysics Letters)}\
  }\textbf {\bibinfo {volume} {123}},\ \bibinfo {pages} {14001} (\bibinfo
  {year} {2018})}\BibitemShut {NoStop}%
\bibitem [{\citenamefont {Taffetani}\ \emph {et~al.}(2019)\citenamefont
  {Taffetani}, \citenamefont {Box}, \citenamefont {Neveu},\ and\ \citenamefont
  {Vella}}]{taffetani2019limitations}%
  \BibitemOpen
  \bibfield  {author} {\bibinfo {author} {\bibfnamefont {Matteo}\ \bibnamefont
  {Taffetani}}, \bibinfo {author} {\bibfnamefont {Finn}\ \bibnamefont {Box}},
  \bibinfo {author} {\bibfnamefont {Arthur}\ \bibnamefont {Neveu}}, \ and\
  \bibinfo {author} {\bibfnamefont {Dominic}\ \bibnamefont {Vella}},\
  }\bibfield  {title} {\enquote {\bibinfo {title} {Limitations of
  curvature-induced rigidity: How a curved strip buckles under gravity},}\
  }\href@noop {} {\bibfield  {journal} {\bibinfo  {journal} {EPL (Europhysics
  Letters)}\ }\textbf {\bibinfo {volume} {127}},\ \bibinfo {pages} {14001}
  (\bibinfo {year} {2019})}\BibitemShut {NoStop}%
\bibitem [{\citenamefont {Audoly}\ and\ \citenamefont
  {Pomeau}(2010)}]{audoly2010elasticity}%
  \BibitemOpen
  \bibfield  {author} {\bibinfo {author} {\bibfnamefont {Basile}\ \bibnamefont
  {Audoly}}\ and\ \bibinfo {author} {\bibfnamefont {Yves}\ \bibnamefont
  {Pomeau}},\ }\bibfield  {title} {\enquote {\bibinfo {title} {Elasticity and
  geometry: from hair curls to the non-linear response of shells},}\
  }\href@noop {} {\  (\bibinfo {year} {2010})}\BibitemShut {NoStop}%
\bibitem [{\citenamefont {Kramer}\ and\ \citenamefont
  {Witten}(1997)}]{kramer1997stress}%
  \BibitemOpen
  \bibfield  {author} {\bibinfo {author} {\bibfnamefont {Eric~M}\ \bibnamefont
  {Kramer}}\ and\ \bibinfo {author} {\bibfnamefont {Thomas~A}\ \bibnamefont
  {Witten}},\ }\bibfield  {title} {\enquote {\bibinfo {title} {Stress
  condensation in crushed elastic manifolds},}\ }\href@noop {} {\bibfield
  {journal} {\bibinfo  {journal} {Physical Review Letters}\ }\textbf {\bibinfo
  {volume} {78}},\ \bibinfo {pages} {1303} (\bibinfo {year}
  {1997})}\BibitemShut {NoStop}%
\bibitem [{\citenamefont {Seung}\ and\ \citenamefont
  {Nelson}(1988)}]{seung1988defects}%
  \BibitemOpen
  \bibfield  {author} {\bibinfo {author} {\bibfnamefont {Hyunjune~Sebastian}\
  \bibnamefont {Seung}}\ and\ \bibinfo {author} {\bibfnamefont {David~R}\
  \bibnamefont {Nelson}},\ }\bibfield  {title} {\enquote {\bibinfo {title}
  {Defects in flexible membranes with crystalline order},}\ }\href@noop {}
  {\bibfield  {journal} {\bibinfo  {journal} {Physical Review A}\ }\textbf
  {\bibinfo {volume} {38}},\ \bibinfo {pages} {1005} (\bibinfo {year}
  {1988})}\BibitemShut {NoStop}%
\bibitem [{\citenamefont {Curtin}\ and\ \citenamefont
  {Scher}(1990)}]{curtin1990mechanics}%
  \BibitemOpen
  \bibfield  {author} {\bibinfo {author} {\bibfnamefont {WA}~\bibnamefont
  {Curtin}}\ and\ \bibinfo {author} {\bibfnamefont {H}~\bibnamefont {Scher}},\
  }\bibfield  {title} {\enquote {\bibinfo {title} {Mechanics modeling using a
  spring network},}\ }\href@noop {} {\bibfield  {journal} {\bibinfo  {journal}
  {Journal of Materials Research}\ }\textbf {\bibinfo {volume} {5}},\ \bibinfo
  {pages} {554--562} (\bibinfo {year} {1990})}\BibitemShut {NoStop}%
\bibitem [{\citenamefont {Ostoja-Starzewski}(2002)}]{ostoja2002lattice}%
  \BibitemOpen
  \bibfield  {author} {\bibinfo {author} {\bibfnamefont {Martin}\ \bibnamefont
  {Ostoja-Starzewski}},\ }\bibfield  {title} {\enquote {\bibinfo {title}
  {Lattice models in micromechanics},}\ }\href@noop {} {\bibfield  {journal}
  {\bibinfo  {journal} {Appl. Mech. Rev.}\ }\textbf {\bibinfo {volume} {55}},\
  \bibinfo {pages} {35--60} (\bibinfo {year} {2002})}\BibitemShut {NoStop}%
\bibitem [{\citenamefont {Omori}\ \emph {et~al.}(2011)\citenamefont {Omori},
  \citenamefont {Ishikawa}, \citenamefont {Barth{\`e}s-Biesel}, \citenamefont
  {Salsac}, \citenamefont {Walter}, \citenamefont {Imai},\ and\ \citenamefont
  {Yamaguchi}}]{omori2011comparison}%
  \BibitemOpen
  \bibfield  {author} {\bibinfo {author} {\bibfnamefont {T}~\bibnamefont
  {Omori}}, \bibinfo {author} {\bibfnamefont {T}~\bibnamefont {Ishikawa}},
  \bibinfo {author} {\bibfnamefont {D}~\bibnamefont {Barth{\`e}s-Biesel}},
  \bibinfo {author} {\bibfnamefont {A-V}\ \bibnamefont {Salsac}}, \bibinfo
  {author} {\bibfnamefont {J}~\bibnamefont {Walter}}, \bibinfo {author}
  {\bibfnamefont {Y}~\bibnamefont {Imai}}, \ and\ \bibinfo {author}
  {\bibfnamefont {T}~\bibnamefont {Yamaguchi}},\ }\bibfield  {title} {\enquote
  {\bibinfo {title} {Comparison between spring network models and continuum
  constitutive laws: Application to the large deformation of a capsule in shear
  flow},}\ }\href@noop {} {\bibfield  {journal} {\bibinfo  {journal} {Physical
  Review E}\ }\textbf {\bibinfo {volume} {83}},\ \bibinfo {pages} {041918}
  (\bibinfo {year} {2011})}\BibitemShut {NoStop}%
\bibitem [{\citenamefont {DiDonna}(2002)}]{didonna2002scaling}%
  \BibitemOpen
  \bibfield  {author} {\bibinfo {author} {\bibfnamefont {Brian~Anthony}\
  \bibnamefont {DiDonna}},\ }\bibfield  {title} {\enquote {\bibinfo {title}
  {Scaling of the buckling transition of ridges in thin sheets},}\ }\href@noop
  {} {\bibfield  {journal} {\bibinfo  {journal} {Physical Review E}\ }\textbf
  {\bibinfo {volume} {66}},\ \bibinfo {pages} {016601} (\bibinfo {year}
  {2002})}\BibitemShut {NoStop}%
\bibitem [{\citenamefont {at~[movie URL] for a representation of the~strain
  map}\ and\ \citenamefont {the sheet profile during a simulation of a sheet
  with~1801x471x2 vertices.}()}]{video}%
  \BibitemOpen
  \bibfield  {author} {\bibinfo {author} {\bibfnamefont {See
  Supplemental~Material}\ \bibnamefont {at~[movie URL] for a representation of
  the~strain map}}\ and\ \bibinfo {author} {\bibnamefont {the sheet profile
  during a simulation of a sheet with~1801x471x2 vertices.}},\ }\href@noop {}
  {}\BibitemShut {NoStop}%
\bibitem [{\citenamefont {Hutchinson}(2010)}]{hutchinson2010knockdown}%
  \BibitemOpen
  \bibfield  {author} {\bibinfo {author} {\bibfnamefont {John~W}\ \bibnamefont
  {Hutchinson}},\ }\bibfield  {title} {\enquote {\bibinfo {title} {Knockdown
  factors for buckling of cylindrical and spherical shells subject to reduced
  biaxial membrane stress},}\ }\href@noop {} {\bibfield  {journal} {\bibinfo
  {journal} {International Journal of Solids and Structures}\ }\textbf
  {\bibinfo {volume} {47}},\ \bibinfo {pages} {1443--1448} (\bibinfo {year}
  {2010})}\BibitemShut {NoStop}%
\bibitem [{\citenamefont {Alava}\ and\ \citenamefont
  {Niskanen}(2006)}]{alava2006physics}%
  \BibitemOpen
  \bibfield  {author} {\bibinfo {author} {\bibfnamefont {Mikko}\ \bibnamefont
  {Alava}}\ and\ \bibinfo {author} {\bibfnamefont {Kaarlo}\ \bibnamefont
  {Niskanen}},\ }\bibfield  {title} {\enquote {\bibinfo {title} {The physics of
  paper},}\ }\href@noop {} {\bibfield  {journal} {\bibinfo  {journal} {Reports
  on progress in physics}\ }\textbf {\bibinfo {volume} {69}},\ \bibinfo {pages}
  {669} (\bibinfo {year} {2006})}\BibitemShut {NoStop}%
\bibitem [{\citenamefont {Peterson}\ \emph {et~al.}(1968)\citenamefont
  {Peterson}, \citenamefont {Seide},\ and\ \citenamefont
  {Weingarten}}]{peterson1968buckling}%
  \BibitemOpen
  \bibfield  {author} {\bibinfo {author} {\bibfnamefont {JP}~\bibnamefont
  {Peterson}}, \bibinfo {author} {\bibfnamefont {P}~\bibnamefont {Seide}}, \
  and\ \bibinfo {author} {\bibfnamefont {VI}~\bibnamefont {Weingarten}},\
  }\bibfield  {title} {\enquote {\bibinfo {title} {Buckling of thin-walled
  circular cylinders},}\ }\href@noop {} {\  (\bibinfo {year}
  {1968})}\BibitemShut {NoStop}%
\bibitem [{\citenamefont {Singer}\ \emph {et~al.}(2002)\citenamefont {Singer},
  \citenamefont {Arbocz},\ and\ \citenamefont {Weller}}]{singer2002vol}%
  \BibitemOpen
  \bibfield  {author} {\bibinfo {author} {\bibfnamefont {J}~\bibnamefont
  {Singer}}, \bibinfo {author} {\bibfnamefont {J}~\bibnamefont {Arbocz}}, \
  and\ \bibinfo {author} {\bibfnamefont {T}~\bibnamefont {Weller}},\ }\bibfield
   {title} {\enquote {\bibinfo {title} {Vol. 2, shells, built-up structures,
  composites and additional topics},}\ }\href@noop {} {\bibfield  {journal}
  {\bibinfo  {journal} {Buckling Experiments: Experimental methods in buckling
  of thin-walled structures}\ } (\bibinfo {year} {2002})}\BibitemShut {NoStop}%
\bibitem [{\citenamefont {Gerasimidis}\ \emph {et~al.}(2018)\citenamefont
  {Gerasimidis}, \citenamefont {Virot}, \citenamefont {Hutchinson},\ and\
  \citenamefont {Rubinstein}}]{gerasimidis2018establishing}%
  \BibitemOpen
  \bibfield  {author} {\bibinfo {author} {\bibfnamefont {S}~\bibnamefont
  {Gerasimidis}}, \bibinfo {author} {\bibfnamefont {E}~\bibnamefont {Virot}},
  \bibinfo {author} {\bibfnamefont {JW}~\bibnamefont {Hutchinson}}, \ and\
  \bibinfo {author} {\bibfnamefont {SM}~\bibnamefont {Rubinstein}},\ }\bibfield
   {title} {\enquote {\bibinfo {title} {On establishing buckling knockdowns for
  imperfection-sensitive shell structures},}\ }\href@noop {} {\bibfield
  {journal} {\bibinfo  {journal} {Journal of Applied Mechanics}\ }\textbf
  {\bibinfo {volume} {85}} (\bibinfo {year} {2018})}\BibitemShut {NoStop}%
\bibitem [{\citenamefont {Virot}\ \emph {et~al.}(2017)\citenamefont {Virot},
  \citenamefont {Kreilos}, \citenamefont {Schneider},\ and\ \citenamefont
  {Rubinstein}}]{virot2017stability}%
  \BibitemOpen
  \bibfield  {author} {\bibinfo {author} {\bibfnamefont {Emmanuel}\
  \bibnamefont {Virot}}, \bibinfo {author} {\bibfnamefont {Tobias}\
  \bibnamefont {Kreilos}}, \bibinfo {author} {\bibfnamefont {Tobias~M}\
  \bibnamefont {Schneider}}, \ and\ \bibinfo {author} {\bibfnamefont
  {Shmuel~M}\ \bibnamefont {Rubinstein}},\ }\bibfield  {title} {\enquote
  {\bibinfo {title} {Stability landscape of shell buckling},}\ }\href@noop {}
  {\bibfield  {journal} {\bibinfo  {journal} {Physical review letters}\
  }\textbf {\bibinfo {volume} {119}},\ \bibinfo {pages} {224101} (\bibinfo
  {year} {2017})}\BibitemShut {NoStop}%
\end{thebibliography}%

\end{document}